\documentclass[11pt,titlepage]{article}
\usepackage{fullpage}

\def\be{\begin{equation}}
\def\ee{\end{equation}}
\def\bea{\begin{eqnarray}}
\def\eea{\end{eqnarray}}
\gdef\makemath#1{\ifmmode #1 \else $ #1 $\fi}
\newcommand{\ignore}[1]{}

\newwrite\noteFile%
\immediate\openout\noteFile=\jobname.notes%

\def\saveNote#1{%
\immediate\write\noteFile{Page \expandafter\thepage; }%
\immediate\write\noteFile{\expandafter#1}%
\immediate\write\noteFile{}%
}

\def\authNote#1{}

\def\QED{\hfill\fbox{\phantom{.}}}

\newtheorem{theorem}{Theorem}[section]
\newtheorem{thm}{Theorem}[section]

\newtheorem{cor}[theorem]{Corollary}
\newtheorem{lemma}[theorem]{Lemma}
\newtheorem{lem}[theorem]{Lemma}

\newtheorem{defn}[theorem]{Definition}

\newtheorem{prop}[theorem]{Proposition}

{\xdef\kindOfTheorem{#1}\begin{\kindOfTheorem}}%
{\end{\kindOfTheorem}}%

{\par\medskip\noindent{\bf #1}\begin{it}}%
{\end{it}\par\medskip}

\newcommand{\ket}[1]{{|{#1} \rangle}}
\newcommand{\bra}[1]{{\langle {#1}|}}

\newcommand{\nums}{{\bf N}}
\newcommand{\nats}{{\nums}}
\newcommand{\ints}{{\bf Z}}
\newcommand{\rats}{{\bf Q}}

\newcommand{\complexes}{{\bf C}}

\newcommand{\QACCP}{QACC$_{pl}^{\log}$}
\newcommand{\QACCG}{QACC$_{gates}^{\log}$}
\newcommand{\BQACCG}{BQACC$_{\rats,gates}^{\log}$}
\newcommand{\NQACCG}{NQACC$_{gates}^{\log}$}
\newcommand{\EQACCG}{EQACC$_{gates}^{\log}$}
\newcommand{\BQACCP}{BQACC$_{\rats,pl}^{\log}$}
\newcommand{\NQACCP}{NQACC$_{pl}^{\log}$}
\newcommand{\EQACCP}{EQACC$_{pl}^{\log}$}

\newcommand{\Mod}{{\rm Mod}}
\newcommand{\MOD}{{\rm MOD}}

\newenvironment{proof}%
{\medskip

\noindent {\bf Proof.}
}
{\QED \medskip}

\newcommand{\monus}{\mathbin{\mathchoice%
{\buildrel .\lower.6ex\hbox{\vphantom{.}} \over {\smash-}}%
{\buildrel .\lower.6ex\hbox{\vphantom{.}} \over {\smash-}}%
{\buildrel .\lower.4ex\hbox{\vphantom{.}} \over {\smash-}}%
{\buildrel .\lower.3ex\hbox{\vphantom{.}} \over {\smash-}}}}

\newcommand{\GN}[1]{\,\!^{\lceil}\!#1\,\!^{\rceil}}

\newcommand{\AND}{\wedge}

\newcommand{\HALF}[1]{\lfloor\frac{1}{2}#1\rfloor}

\begingroup\makeatletter\ifx\SetFigFont\undefined
\def\x#1#2#3#4#5#6#7\relax{\def\x{#1#2#3#4#5#6}}%
\expandafter\x\fmtname xxxxxx\relax \def\y{splain}%
\ifx\x\y   % LaTeX or SliTeX?
\gdef\SetFigFont#1#2#3{%
  \ifnum #1<17\tiny\else \ifnum #1<20\small\else
  \ifnum #1<24\normalsize\else \ifnum #1<29\large\else
  \ifnum #1<34\Large\else \ifnum #1<41\LARGE\else
     \huge\fi\fi\fi\fi\fi\fi
  \csname #3\endcsname}%
\else
\gdef\SetFigFont#1#2#3{\begingroup
  \count@#1\relax \ifnum 25<\count@\count@25\fi
  \def\x{\endgroup\@setsize\SetFigFont{#2pt}}%
  \expandafter\x
    \csname \romannumeral\the\count@ pt\expandafter\endcsname
    \csname @\romannumeral\the\count@ pt\endcsname
  \csname #3\endcsname}%
\fi
\fi\endgroup

\begin{document}

%\date{{\em \today -- Draft}}
\title{On the Complexity of Quantum ACC}
\author{
Frederic Green\\
Department of Mathematics and Computer Science\\
Clark University, Worcester, MA 01610\\
fgreen@black.clarku.edu\\
\and
Steven Homer\\
Computer Science Department\\ 
Boston University, Boston, MA 02215\\
homer@cs.bu.edu\\
\and
Christopher Pollett\\
Department of Mathematics\\
University of California, Los Angeles, CA\\
cpollett@willow.math.ucla.edu\\
}
\maketitle

\thispagestyle{empty}

\begin{abstract}
%  We show that, for any $q > 1$, a quantum gate that determines if the
%number of 1's in the input is divisible by $q$ is equivalent (up to constant
%depth) to any quantum gate that determines divisibility by another
%number, and is in turn equivalent to the ability to fan out inputs.
  For any $q > 1$, let $\MOD_q$ be a quantum gate that determines if the
number of 1's in the input is divisible by $q$.
We show that for any $q,t > 1$, $\MOD_q$ is equivalent to $\MOD_t$
(up to constant depth). 
Based on the case $q=2$, Moore~\cite{moore99} has 
shown that quantum analogs of 
AC$^{(0)}$, ACC$[q]$, and ACC,
denoted QAC$^{(0)}_{wf}$, QACC$[2]$, QACC  respectively,
define the same class of operators, leaving $q > 2$ as an
open question. Our result resolves this question, proving that
QAC$^{(0)}_{wf} =$ QACC$[q] = $ QACC for all $q$.
We also develop techniques for proving upper bounds for QACC in terms of related
language classes. We 
define classes of languages EQACC, NQACC and BQACC$_{\rats}$. We define a notion of 
$\log$-planar QACC operators
and show the appropriately restricted versions of EQACC and NQACC
are contained in P/poly. We also define a notion of $\log$-gate restricted
QACC operators and show the appropriately restricted versions
of EQACC and NQACC are contained in
TC$^{(0)}$.  To do this last proof, we show that TC$^{(0)}$
can perform iterated addition and multiplication in certain field
extensions.  We also introduce the notion of a polynomial-size tensor graph and we
show that families of such graphs can encode the amplitudes resulting from applying
an arbitrary QACC operator
to an initial state.
%Looking at what information one can extract from
%these graphs under various complexity limitations gives the TC$^{(0)}$ and
%$P/poly$ result.
\end{abstract}

\section{Introduction}%%of quaccrev.tex

Advances in quantum computation in the last decade have been
among the most notable in theoretical computer
science. This is due to the surprising improvements in 
the efficiency of solving several fundamental  combinatorial problems
using quantum mechanical methods in place  of their classical counterparts.
These advances led to considerable efforts in finding 
new efficient
quantum algorithms for 
classical problems and  in developing a complexity
theory of quantum computation. 
                                         
While most of the original results in quantum computation were 
developed using quantum Turing machines, 
they can also be formulated in terms of quantum circuits, which
yield a more natural model of quantum
computation. For example, Shor
\cite{shor97} has shown that quantum circuits can  factor integers more 
efficiently 
than
any known classical algorithm for factoring.  And quantum circuits have 
been shown 
(see Yao
\cite{yao93}) to provide a universal model for quantum computation. 

%    A fruitful approach to understanding the
%power of quantum computers is to study quantum analogs of
%classical complexity classes. Usually, quantum complexity classes
%have higher (classical) complexity than their classical
%counterparts. Often, we can also obtain non-trivial upper bounds on the
%classical complexity of quantum complexity classes. This paper is a
%contribution to this line of research in the realm of circuit complexity.

In the classical setting, small depth circuits are considered a 
good model for parallel computing. Constant-depth circuits, corresponding
to constant parallel time, are of central importance. For
example, constant-depth circuits of AND, OR and NOT gates of polynomial 
size
(called AC$^{(0)}$ circuits) can add and subtract binary numbers. The 
class ACC
extends AC$^{(0)}$ by allowing modular counting gates. The class 
TC$^{(0)}$,
consisting of constant-depth threshold circuits, can compute iterated
multiplication.

In studying quantum circuits, it is natural to consider the power of 
small depth circuit families. 
Quantum circuit
models analogous to
the central classical circuit classes have recently  been studied by Moore 
and
Nilsson~\cite{moore98} and Moore~\cite{moore99}. They investigated the
properties of classes of quantum operators
QAC$^{(0)}_{wf}$, QACC$[q]$, and QNC defined to be analogous to and to 
contain
their classical counterparts.   
This paper is a
contribution to this line of research.

For example, a quantum analog of AC$^{(0)}$, defined by
Moore and denoted
QAC$^{(0)}_{wf}$, is the class of families of operators which
can be built out of products of constantly many layers 
consisting of polynomial-sized tensor
products of one-qubit gates (analogous to NOT's), Toffoli gates (analogous
to AND's and OR's) and
fan-out gates\footnote{The subscript ``$wf$" in the notation denotes ``with
fan-out." The idea of fan-out in the quantum setting is subtle, as will be
made clearer later in this paper. See
Moore~\cite{moore99} for a more in-depth discussion.}.
An analog of ACC$[q]$ (i.e., ACC circuit families only allowing
Mod$_q$ gates) is 
QACC$[q]$, defined similarly to QAC$^{(0)}_{wf}$, but
replacing the fan-out gates with
quantum $\Mod_q$ gates (which we denote as $\MOD_q$). QACC
is the same class but we allow $\MOD_q$ gates for every $q$. 
Moore~\cite{moore99} proves the surprising result 
QAC$^{(0)}_{wf}$$ = $ QACC$[2]$ $ = $ QACC. This is in sharp contrast to
the classical result of Smolensky~\cite{smo87} that says ACC$^{(0)}[q] 
\not = $
ACC$^{(0)}[p]$ for any pair of distinct primes $q, p$, which implies
that for any prime $p$, AC$^{(0)} \subset $ ACC$^{(0)}[p] \subset$
ACC. 
This result showed that parity gates are as powerful as any other
mod gates in QACC, but  left open the complexity of $\MOD_q$ gates for
$q > 2$. 
 
 In \cite{moore99}, Moore conjectured that QACC $\not = $ QACC$[q]$ for
odd $q$. In this paper, we
provide the missing ingredients to show that in fact QACC$ =$ QACC$[q]$
for any
$q
\ge 2$. Moore's result showed that parity is as good as any other $\MOD_q$ 
gate;
our result further shows that any $\MOD_q$ gate is as good as any other.
The main technical contribution is the application of the Quantum Fourier
Transform (using complex $q^{th}$ roots of unity), and encodings of
base $q$ digits using qubits.
%This result strengthens Moore's lower bound, in the sense that
%it shows that any quantum $\MOD_q$ gate is powerful enough to capture
%all of QACC (and therefore ACC) in constant depth.
% Does anyone want to add more details here?  

  We also develop methods for proving upper bounds for language classes related
to QACC. Our methods result in upper bounds for restricted QACC circuits. Roughly
speaking, we show that QACC is no more powerful than P/Poly provided that a layer of
``wire-crossings" in the QACC operator can be written as log many compositions of
Kronecker  products of  controlled-not gates. We call this class \QACCP, 
where the ``pl'' is for this planarity condition. We show if one further
restricts attention to the case where the number of multi-line gates 
(gates whose input is more than 1 qubit) is
log-bounded then the circuits are no
more powerful than TC$^{(0)}$. We call this class \QACCG.
These results hold for arbitrary complex amplitudes in the QACC circuits.

To be more precise, it is necessary
to show how a class of operators in QACC can define a
language, as usually
considered in complexity theory. 
In this paper, we
define classes of languages EQACC, NQACC, and 
BQACC based on the expectation of observing a certain state after
applying the QACC operator to the input state. For example, the class NQACC
corresponds to the case where $x$ is in the language if the expectation of the observed
state after applying the QACC operator is non-zero. This is analogous to 
the definition of 
the class NQP in Fenner et 
al.~\cite{fghr98}. 

In this paper, we show that \NQACCG is in 
TC$^{(0)}$ and \NQACCP is in P/poly. Although the proof uses some of the 
techniques developed by 
Yamakami and
Yao~\cite{yy98} to show that  NQP$_{C}=$ co-C$_=$P,
the small depth circuit case presents technical challenges
not present in their setting. In particular, given a QACC operator built 
out
of layers $M_1, \ldots , M_t$ and an input state $\ket{x, 0^{p(n)}}$, we 
must show that a TC$^{(0)}$ circuit can keep track of the
amplitudes of each possible resulting state as each layer is applied. 
After all layers have been applied, the TC$^{(0)}$ circuit then needs to be
able to check that the amplitude of one possible state is
non-zero. Unfortunately, there
could be exponentially many states with non-zero amplitudes after applying 
a 
layer. 
To handle this problem we introduce the idea of a
``tensor-graph," a new way to 
represent a collection of states. We can extract from these graphs 
(via  TC$^{(0)}$ or P/poly computations) 
whether the amplitude of any particular vector 
is 
non-zero.

The exponential growth in the number of states is one of the
primary obstacles to proving that all of NQACC is in TC$^{(0)}$
(or even P/Poly), and thus the tensor graph formalism represents
a significant step
towards such an upper bound. The reason the bounds apply only
in the restricted cases is that although tensor graphs
can represent any QACC operator, in the case of
operators with layers that might do arbitrary permutations, 
the top-down approach we use to compute a desired amplitude from the graph 
no longer seems to work. We feel that it is likely 
that the amplitude of any vector in a tensor graph can be written
as a polynomial product of a polynomial sum in some extension algebra
of the ones we work with in this paper, in which case it is quite
likely it can be evaluated in TC$^{(0)}$.
 
Another important obstacle to obtaining a TC$^{(0)}$ upper bound is
that one needs to
be able to add and multiply a
polynomial number of complex amplitudes that may appear in a QACC computation.
We solve this problem.
It reduces to adding and multiplying polynomially many
elements of a
 certain transcendental extension of the rational numbers. We
show that in fact
 TC$^{(0)}$ is closed under iterated addition and multiplication
of such numbers (Lemma~\ref{sumprod} below). This result is of independent 
interest, 
and our application of tensor-graphs and these closure properties of 
TC$^{(0)}$ may prove useful in further 
investigations of small-depth quantum circuits.

We now discuss the organization of the rest of this paper. In the
next section we introduce the definitions and notations we use in this
paper. Then in the following section we prove QACC$[q]$ $=$ QACC.
Finally, in the last section, we prove the TC$^{(0)}$ and P/poly
upper bounds for the restricted classes discussed above.
% EQACC, NQACC, and
%BQACC$_{\rats}$  are all contained in
%p-uniform
%TC$^{(0)}$.
%Henceforth,
%when we refer to TC$^{(0)}$ we will mean $p$-uniform TC$^{(0)}$.  

\section{Preliminaries}
In this section we define the gates used as building blocks for
our quantum circuits. Classes of operators  built out of these gates are 
then
defined. We define language classes that can be determined by
these operators and give a couple definitions from algebra. Lastly, some
closure properties of TC$^{(0)}$ are described.

\begin{defn}
\item By a {\em one-qubit gate} we mean an operator from the group $U(2)$.

\item Let $U = \left( \begin{array}{cc} u_{00} & u_{01} \\ u_{10} & u_{11}
\end{array}
\right) \in U(2)$. $\AND_m(U)$ is defined as: 
$\AND_0(U)=U$ and for $m>0$, $\AND_m(U)$ is
$$\AND_m(U)(\ket{\vec{x},y}) = \left\{
\begin{array}{ll}
u_{y0}\ket{\vec{x},0} + u_{y1}\ket{\vec{x},1} & \mbox{if }
\AND^m_{k=1}x_k=1\\
\ket{\vec{x},y} & \mbox{otherwise}
\end{array} \right.$$

\item Let $X = \left( \begin{array}{cc} 0 & 1 \\ 1 & 0\end{array} \right)$.
A {\em Tofolli gate} is a $\AND_m(X)$ gate for some $m \geq 0$. 
A {\em controlled-not} gate is a $\AND_1(X)$ gate.

\item An {\em (m-)spaced controlled-not gate} is an operator that
maps $\ket{y_1,\ldots, y_m,x}$ to $\ket{x\oplus y_1, y_2\ldots, y_m,x}$
or $\ket{y_1,\ldots, y_m,x}$ to $\ket{x, y_1\ldots, y_{m-1},y_m\oplus x}$  

\item An {\em  (m-ary) fan out gate} $F$ is an operator that maps from
$\ket{y_1,\ldots, y_m,x}$ to $\ket{x\oplus y_1,\ldots, x\oplus y_m,x}$.

\item A $\MOD_{q,r}$ gate is an operator that maps
$\ket{y_1,\ldots, y_m,x}$ to $\ket{y_1,\ldots, y_m,x \oplus (\sum y_i 
\bmod 
q 
\equiv r) }$.
\end{defn}

We use the following graphical notation for parity
(i.e., $\MOD_2$) or, in the case of $n=1$, for
controlled-not:

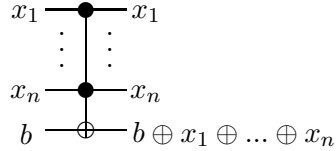
\begin{center}
\begin{picture}(1700,1100)
\put(300,950){\line(1,0){500}}
\put(550.25,950.25){\circle*{100}}
\put(550.25,450.25){\circle*{100}}
\put(550.25,200.25){\circle{100}}
\put(300,450){\line(1,0){500}}
\put(300,200){\line(1,0){500}}
\put(550,950){\line(0,-1){800}}
\put(100,850){\makebox(150,150)[]{$x_1$}}
\put(850,850){\makebox(150,150)[]{$x_1$}}
\put(100,350){\makebox(150,150)[]{$x_n$}}
\put(850,350){\makebox(150,150)[]{$x_n$}}
\put(100,100){\makebox(150,150)[]{$b$}}
\put(1150,100){\makebox(650,150)[]{$b\oplus x_1 \oplus ...\oplus x_n$}}
\put(350,750){\makebox(100,100)[]{.}}
\put(350,650){\makebox(100,100)[]{.}}
\put(350,550){\makebox(100,100)[]{.}}
\put(650,750){\makebox(100,100)[]{.}}
\put(650,650){\makebox(100,100)[]{.}}
\put(650,550){\makebox(100,100)[]{.}}
\end{picture}
\end{center}

\noindent and for $\MOD_q$:

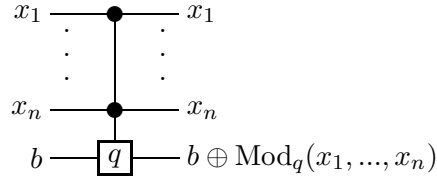
\begin{center}
\begin{picture}(2400,1550)
\put(750,500){\line(-1,0){300}}
\put(950,500){\line(1,0){300}}
\put(450,800){\line(1,0){800}}
\put(450,1400){\line(1,0){800}}
\put(850,1400){\line(0,-1){800}}
\put(850.25,800.25){\circle*{100}}
\put(850.25,1400.25){\circle*{100}}
\put(150,1350){\makebox(250,100)[r]{$x_1$}}
\put(750,400){\framebox(200,200)[]{$q$}}
\put(100,750){\makebox(300,100)[r]{$x_n$}}
\put(1300,1350){\makebox(300,100)[l]{$x_1$}}
\put(1300,750){\makebox(300,100)[l]{$x_n$}}
\put(100,450){\makebox(300,100)[r]{$b$}}
\put(1300,450){\makebox(1000,100)[l]{$b\oplus {\rm Mod}_q(x_1,...,x_n)$}}
\put(500,1250){\makebox(100,100)[]{.}}
\put(500,1100){\makebox(100,100)[]{.}}
\put(500,950){\makebox(100,100)[]{.}}
\put(1100,1250){\makebox(100,100)[]{.}}
\put(1100,1100){\makebox(100,100)[]{.}}
\put(1100,950){\makebox(100,100)[]{.}}
%\put(150,100){\makebox(1400,200)[]{Fig. 2: A MOD$_q$ gate.}}
\end{picture}
\end{center}

  As discussed in~\cite{moore99}, the no-cloning theorem of quantum
mechanics makes it difficult to directly fan out qubits in constant
depth (although constant fan-out is no problem). Thus
it is necessary to define the operator $F$ as in the above definition;
refer to~\cite{moore99} for further details. Also, in the literature
it is frequently the case that one says a given operator $M$ on 
$\ket{y_1, \ldots, y_m}$ can be written as a tensor product of 
certain
gates. What is meant is that there is an permutation operator $\Pi$ ( 
a
map $\ket{y_1, \ldots, y_m}$ to $\ket{y_{\pi(1)},\ldots, y_{\pi(m)}}$ for
some permutation $\pi$) such that 
$$M\ket{y_1, \ldots y_m} = \Pi\otimes^n_j M_j \Pi^{-1}\ket{y_1, \ldots 
y_m}$$
where  the $M_i$'s are our base gates,  i.e., those gates for which no
inherent ordering on the $y_i$ is assumed {\it a priori}. Since it is important 
to
keep track of such details in our upper bounds proofs, we will always use
Kronecker products of the form $\otimes^n_j M_j$ without unspoken 
permutations.
Nevertheless, being able to do permutation operators (not conjugation by a
permutation) intuitively allows our circuits to simulate classical
wire crossings. To handle permutations, we allow our circuits to have
controlled-not layers. A {\em controlled-not layer} is a gate which 
performs,
in one step, controlled-not's between an arbitrary
collection of disjoint pairs of lines in its domain. That is, it performs $\Pi
\otimes^n_j \AND_1(X) \Pi^{-1}$ for some permutation operator $\Pi$. Moore
Nilsson~\cite{moore98} show that any permutation can be written as a finite
product of controlled-not layers. We say a controlled-not layer is
{\em log-depth} if it can be written as the composition of log many
matrices each of which is the Kronecker product of identities and spaced 
controlled-not gates.

$M^{\otimes n}$ is the $n$-fold Kronecker product of $M$ with itself. The
next definitions are based on Moore~\cite{moore99}. 
\begin{defn} 
\item QAC$^{(k)}$ is the class of families $\{F_n\}$, where $F_n$ is in
$U(2^{n+p(n)})$, $p$ a polynomial, and each $F_n$ is writable as a product
of $O(\log^k n)$  layers,  where a {\em layer} is a Kronecker product of
one-qubit gates and Toffoli gates or is a controlled-not layer. Also for
all $n$  the number of distinct types of one qubit gates used must be fixed. 

\item QACC$^{(k)}[q]$ is the same as QAC$^{(k)}$ except we also allow $\MOD_q$ gates. 
QACC$^{(k)} = \cup_q $QACC$^{(k)}[q]$.

\item QAC$^{(k)}_{wf}$ is the same as QAC$^{(k)}$ but we also allow fan-out
gates.

\item QACC is defined as QACC$^{(0)}$ and QACC$[q]$ is defined as
QACC$^{(0)}[q]$.
 \QACCP is QACC restricted to log-depth controlled not
layers.
 \QACCG is QACC restricted so that the total number of
multi-line gates in all layers is log-bounded. 

\item If $\mathcal{C}$ is one of the above classes, then $\mathcal{C}_K$ 
are the families in $\mathcal{C}$ with coefficients restricted to $K$. 

\item Let $\{F_n\}$ and $\{G_n\}$, $G_n, F_n\in U(2^{n})$ be families of
operators. We say $\{F_n\}$ is {\rm QAC$^{(0)}$ reducible}
to $\{G_n\}$ if there is a family
$\{R_n\}$, $R_n\in U(2^{n+p(n)})$ of QAC$^{(0)}$ operators augmented with
operators from $\{G_n\}$ such that for all $n$, ${\bf x},{\bf y}
\in \{0,1\}^n$, there is a setting of $z_1,...,z_{p(n)} \in \{0,1\}$
for which $\bra{{\bf y}}F_n\ket{{\bf x}} = \bra{{\bf y}, {\bf z}}R_n
\ket{{\bf x},{\bf z}}$. Operator families are {\rm QAC$^{(0)}$ equivalent}
if they are QAC$^{(0)}$ reducible to each other. If ${\cal C}_1$
and ${\cal C}_2$ are families of QAC$^{(0)}$ equivalent operators, we write
${\cal C}_1 = {\cal C}_2$.
\end{defn}

%  We also allow, in any family of operators, a polynomial number of
%{\it auxiliary} inputs (called ``ancillae" in~\cite{moore99}), which are 
%set
%to constant values (usually 0) 
%at the beginning of a computation, and which must be returned to their
%original values at the end.

We refer to the $z_i$'s above as
 ``auxiliary bits" (called ``ancillae" in \cite{moore99}).
Note that in proving QAC$^{(0)}$ equivalence, the auxiliary bits
must be returned to
their original values in a  computation.

It follows for any $\{F_n\}\in$ QAC$^{(0)}$ that $F_n$ is writable as a 
product 
of finite number of layers. Moore~\cite{moore99} shows
QAC$^{(0)}_{wf} =$ QACC$[2]$ $=$ QACC. Moore~\cite{moore99} places
no restriction on  the number of  distinct types of one-qubit gates used 
in a
given family of operators. We do this so that the number of distinct
amplitudes which appear in matrices in a layer is fixed with respect to 
$n$.
This restriction arises implicitly  in the quantum Turing machine case of 
the
upper bounds proofs in Fenner, et al.~\cite{fghr98} and Yamakami and 
Yao~\cite{yy98}. Also, it seems fairly natural since in the classical case 
one
builds circuits  using a fixed number of distinct gate types. Our classes 
are,
thus, more  ``uniform'' than Moore's. We now
define language classes based on our classes of operator families.

\begin{defn}
Let $\mathcal{C}$ be a class of families of $U(2^{n+p(n)})$ operators 
where 
$p$ is
 a polynomial and $n=|x|$.
\begin{enumerate}
\item E$\cdot\mathcal{C}$ is the class of languages $L$ such that for some 
$\{F_n\} \in \mathcal{C}$ and 
$\{\bra{\vec{z}_n}\}=\{\bra{z_{n,1}, \ldots, z_{n,n+p(n)}}\}$ a family
of states,
$m:=|\bra{\vec{z}_n}F_n\ket{x, 0^{p(n)}}|^2$ is $1$ or $0$ and $x\in L$ 
iff 
$m=1$.  
\item N$\cdot\mathcal{C}$ is the class of languages $L$ such that for some 
$\{F_n\} \in \mathcal{C}$ and $\{\bra{\vec{z}_n}\}$ a family of states,
$x\in L$ iff $|\bra{\vec{z}_n}F_n\ket{x, 0^{p(n)}}|^2>0$.
\item B$\cdot\mathcal{C}$ is the class of languages $L$ so that for some 
$\{F_n\} \in \mathcal{C}$ and $\{\bra{\vec{z}}\}$,
$x\in L$ if $|\bra{\vec{z}_n}F_n\ket{x, 0^{p(n)}}|^2>3/4$ and $x\not\in 
L$ 
if $|\bra{\vec{z}_n}F_n\ket{x, 0^{p(n)}}|^2<1/4$ .  
\end{enumerate}
\end{defn}

It follows E$\cdot\mathcal{C}\subseteq$ N$\cdot\mathcal{C}$
and E$\cdot\mathcal{C}\subseteq$ B$\cdot\mathcal{C}$. We frequently will 
omit
the `$\cdot$' when writing a class, so E$\cdot$QACC is written as 
EQACC. Let $\ket{\Psi} := F_n\ket{x, 0^{p(n)}}$.
Notice that
$|\bra{\vec{z}_n}F_n\ket{x, 0^{p(n)}}|^2 = 
\bra{\Psi}P_{\ket{\vec{z}_n}}\ket{\Psi}$,
where $P_{\ket{\vec{z}_n}}$ is the projection matrix onto 
$\ket{\vec{z}_n}$. We could allow in our definitions measurements
of up to polynomially many such projection observables and not affect 
our results below. 
However, this would shift the burden of the computation in some sense away 
from
the QACC operator and instead onto preparation of the observable.

Next are some variations on familiar definitions from algebra.

\begin{defn}
Let $k>0$. A subset $\{\beta_i\}_{1\leq i\leq k}$ of $\complexes$ is {\em 
linearly independent} if $\sum^k_{i=1} a_i\beta_i \neq 0$ for any  
$(a_1,\ldots, a_k) \in \rats^k - \{\vec{0}^k\}$.
A set $\{\beta_i\}_{1\leq i\leq k}$ is {\em algebraically independent} 
if the only $p\in\rats[x_1,\ldots,x_k]$ with $p(\beta_1,\ldots,\beta_k)=0$ 
is
the zero polynomial. \end{defn}

We now briefly mention some closure properties of TC$^{(0)}$ computable
functions that are useful in proving  \NQACCG$ \subseteq$ TC$^{(0)}$.
For proofs of the statements
in the next lemma see~\cite{siu91,siu94,clote93}.

\begin{lemma}
(1) TC$^{(0)}$ functions are closed under composition. (2) The following 
are
TC$^{(0)}$ computable: $x+y$, $x\monus y:=  x - y$ if $x-y >0$ and $0$
otherwise, $|x| := \lceil \log_2(x + 1 )\rceil$, $x\cdot y$, $\lfloor x/y
\rfloor$, $2^{\min(i,p(|x|)}$, and $cond(x,y,z) := y$ if $x>0$ and $z$
otherwise. (3) If $f(i,x)$ is a TC$^{(0)}$ computable then
$\sum^{p(|x|)}_{k=0} f(k,x)$, $\prod^{p(|x|)}_{k=0} f(k,x)$, $\forall i 
\leq
p(|x|)(f(i,x)=0)$, $\exists i \leq p(|x|)(f(i,x)=0)$, and $\mu i \leq
p(|x|)(f(i,x)=0):=$ the least $i$ such that $f(i,x)=0$ or $p(x)+1$ 
otherwise,
are TC$^{(0)}$ computable.  
\end{lemma}
We drop the $\min$ from the $2^{\min(i,p(|x|))}$ when it is obvious a 
suitably
large $p(|x|)$ can be found. We define $max(x,y):= cond(1\monus (y\monus
x)),x,y)$ and define
\begin{eqnarray*}
max_{i\leq p(|x|)}(f(i)) & := &\\
 (\mu i  \leq  &p(|x|)&)(\forall j \leq
p(|x|)(f(j)\monus f(i) = 0)
\end{eqnarray*}
Using the above functions we describe a way to do sequence coding
in TC$^{(0)}$. Let $\beta_{|t|}(x,w) := \lfloor(w \monus  \lfloor
w/2^{(x+1)|t|}\rfloor\cdot2^{(x+1)|t|})/2^{x|t|}\rfloor.$ The function 
$\beta_{|t|}$ is useful for block coding. Roughly, $\beta_{|t|}$
first gets rid of the bits after the $(x+1)|t|$th bit then chops off the 
low order $x|t|$ bits. Let $B=
2^{|\max(x,y)|}$, so that $B$ is longer than either $x$ or $y$.  Hence, we
code pairs as $\langle x,y \rangle := (B+y)\cdot 2B + B+x$, and
projections as $(w)_1 := 
\beta_{\HALF{|w|}\monus 1}(0, \beta_{\HALF{|w|}}(0,w))$ and   $(w)_2 :=
\beta_{\HALF{|w|}\monus 1}(0,\beta_{\HALF{|w|}}(1,w))$. We can encode a
poly-length, TC$^{(0)}$ computable sequence of numbers  $\langle f(1),
\ldots, f(k) \rangle$ as the pair  $\langle \sum^k_i(f(i)2^{i\cdot m}),
m\rangle$ where $m:=|f(\max_i(f(i)))|+1$. We then define the function
which projects out the $i$th member of a sequence as  $\beta(i,w) :=
\beta_{(w)_2}(i,w)$. 

We can code integers using the positive natural numbers by letting the 
negative 
integers be the odd natural numbers and the positive
integers be the even natural numbers. TC$^{(0)}$ can use the
TC$^{(0)}$ circuits for natural numbers to compute both the polynomial sum 
and polynomial product
of a sequence of TC$^{(0)}$ definable integers. It can also compute the
rounded quotient of two such integers. For instance, to do a polynomial
sum of integers, compute the natural number which is the sum of the 
positive numbers in the sum using $cond$ and our natural number
iterated addition circuit. Then compute the natural number which is the sum
of the negative numbers in the sum. Use the subtraction
circuit  to subtract the smaller from the larger number and multiply by 
two.
One is  then added if the number should be negative. For products, we 
compute
the  product of the natural numbers which results by dividing each integer
code by two and  rounding down. We multiply the result by two. We then sum 
the
number of terms in our  product which were negative integers. If this 
number is
odd we add one to  the product we just calculated. Finally, division can be
computed using the Taylor expansion of $1/x$.

\section{QACC[$q$]}
In this section, we show QACC[$q$]$=$QACC for any $q\ge 2$.
 
  Let $q \in \nums$, $q \ge 2$ be fixed throughout this discussion. Consider
quantum states labelled by digits in $D = \{0,...,q-1\}$. By analogy with
``qubit," we refer to a state of the form, $$\sum\limits_{k=0}^{q-1}
c_k\ket{k}$$ with $\sum_k |c_k|^2 = 1$ as a ``qudigit." Direct products of the
basis states will be labelled by lists of eigenvalues, e.g., $\ket{x}\ket{y}$ is
denoted as $\ket{x,y}$.

We define three important operations on qudigits. The $n$-ary {\it modular
addition} operator $M_q$ acts as follows:
$$ M_q \ket{x_1,...,x_n, b} = \ket{x_1,...x_n,
(b + x_1 + ... + x_n)\bmod q}$$
The gate is represented graphically as in the following figure:
%\begin{figure}[h]
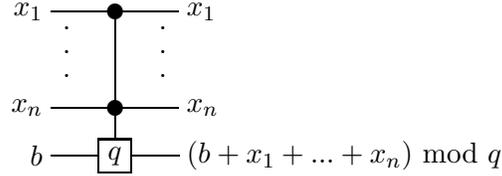
\begin{center}
\begin{picture}(2400,1550)
\put(750,500){\line(-1,0){300}}
\put(950,500){\line(1,0){300}}
\put(450,800){\line(1,0){800}}
\put(450,1400){\line(1,0){800}}
\put(850,1400){\line(0,-1){800}}
\put(850.25,800.25){\circle*{100}}
\put(850.25,1400.25){\circle*{100}}
\put(150,1350){\makebox(250,100)[r]{$x_1$}}
\put(750,400){\framebox(200,200)[]{$q$}}
\put(100,750){\makebox(300,100)[r]{$x_n$}}
\put(1300,1350){\makebox(300,100)[l]{$x_1$}}
\put(1300,750){\makebox(300,100)[l]{$x_n$}}
\put(100,450){\makebox(300,100)[r]{$b$}}
\put(1300,450){\makebox(1000,100)[l]{$(b + x_1 + ... + x_n)~{\rm mod}~q$}}
\put(500,1250){\makebox(100,100)[]{.}}
\put(500,1100){\makebox(100,100)[]{.}}
\put(500,950){\makebox(100,100)[]{.}}
\put(1100,1250){\makebox(100,100)[]{.}}
\put(1100,1100){\makebox(100,100)[]{.}}
\put(1100,950){\makebox(100,100)[]{.}}
%\put(150,100){\makebox(1400,200)[]{Fig. 1: An $M_q$ gate.}}
\end{picture}
\end{center}
%  \caption{An $M_q$ gate.}
%\end{figure}

Since $M_q$ merely permutes the states, it is clear that it is unitary.
Similarly, the $n$-ary unitary {\it base $q$ fanout} operator $F_q$ acts as,
$$ F_q \ket{x_1,...x_n,b} = \ket{(x_1 + b)\bmod q,...(x_n + b)\bmod q,b}.$$
We write $F$ for $F_2$, since it is the ``standard" fan-out gate introduced
by
 Moore (see Definition 2.1).
Note that $M^{-1}_q = M_q^{q-1}$ and $F^{-1}_q = F_q^{q-1}$.

Finally, the Quantum Fourier Transform $H_q$ (which generalizes the Hadamard
transform $H$ on qubits) acts on a single qudigit as,
$$ H_q \ket{a} = {1\over \sqrt{q}}\sum\limits_{b=0}^{q-1} \zeta^{ab} \ket{b},$$
where $\zeta = e^{2\pi i \over q}$ is a primitive complex $q^{th}$ root of 
unity. It is easy to see that $H_q$ is unitary, via the fact that
$\sum_{\ell = 0}^{q-1} \zeta^{a\ell} = 0$ iff $ a \not \equiv 0\bmod q$.

 The first observation is that, analogous to parity and fanout for Boolean
inputs, the operators $M_q$ and $F_q$ are ``conjugates" in the following sense.
%(In the below, $A^{\dag}$ denotes the Hermitian conjugate, or complex
%conjugate transpose, of $A$.)

\begin{prop}\label{conjugates}
  $M_q = (H_q^{\otimes (n+1)})^{-1} F^{-1}_q H_q^{\otimes (n+1)}.$
\end{prop}
\begin{proof}
  We apply the operators $H_q^{\otimes (n+1)}$, $F^{-1}_q$, and
$(H_q^{\otimes (n+1)})^{-1}$ in that order to the state $\ket{x_1,...,x_n, b}$,
and check that the result has the same effect as $M_q$.

The operator $H_q^{\otimes (n+1)}$ simply applies $H_q$ to
each of the $n+1$ qudigits of $\ket{x_1,...,x_n, b}$, which yields,
\begin{eqnarray*}
         {1 \over q^{(n+1)\over 2}}\sum\limits_{{\bf y} \in D^n}
          \sum\limits_{a = 0}^{q-1} \zeta^{{\bf x}\cdot {\bf y} + ab}
               \ket{y_1,...,y_n, a},
\end{eqnarray*}
where ${\bf y}$ is a compact notation for $y_1,...,y_n$, and
${\bf x}\cdot{\bf y}$ denotes $\sum_{i=1}^n x_iy_i$.
Then applying $F^{-1}_q$ to the above state yields,
\begin{eqnarray*}
     {1 \over q^{(n+1)\over 2}}\sum\limits_{{\bf y} \in D^n}
          \sum\limits_{a = 0}^{q-1}&\zeta&^{{\bf x}\cdot {\bf y} + ab}\\
               |(&y_1&-a)\bmod q,...,(y_n - a)\bmod q, a\rangle.
\end{eqnarray*}
By a change of variable, the above can be re-written as,
\begin{eqnarray*}
     {1 \over q^{(n+1)\over 2}}\sum\limits_{{\bf y} \in D^n}
          \sum\limits_{a = 0}^{q-1} \zeta^{\sum_{i=1}^n x_i(y_i+a) + ab}
               \ket{y_1,...,y_n, a}
\end{eqnarray*}
Finally, applying $(H_q^{\otimes (n+1)})^{-1}$ to the above undoes the
Fourier transform and
puts the coefficient of $a$ in the exponent into the last slot of the state.
The result is,
\begin{eqnarray*}
  (H_q^{\otimes (n+1)})^{-1} F^{-1}_q H_q^{\otimes (n+1)}
  \ket{x_1,...,x_n,b} =\\
             \ket{x_1,...,x_n, (b + x_1 + ... + x_n) \bmod q},
\end{eqnarray*}
which is exactly what $M_q$ would yield.

\end{proof}

  We now describe how the operators $M_q$, $F_q$ and $H_q$ can be modified to
operate on registers consisting of qu{\it bits} rather than qu{\it digits}.
Firstly, we encode each digit using $\lceil \log q \rceil$ bits. Thus, for
example, when $q = 3$, the basis states $\ket{0}, \ket {1}$ and $\ket{2}$ are
represented by the two-qubit registers $\ket{00}, \ket{01}$ and $\ket{10}$,
respectively. Note that there remains one state (in the example, $\ket{11}$)
which does not correspond to any of the qudigits. In general, there will be
$2^{\lceil \log q\rceil} - q$ such ``non-qudigit" states. $M_q$, $F_q$ and
$H_q$ can now be defined to act on qubit registers, as follows. Consider a
state $\ket{x}$ where $x$ is a number represented as $m$ bits (i.e., an
$m$-qubit register). If $m < \lceil \log q\rceil$, then $H_q$ leaves $\ket{x}$
unaffected. If $0 \le x \le q -1$ (where here we are identifying $x$ with the
number it represents), then $H_q$ acts exactly as one expects, namely,
$ H_q\ket{x} = (1 / \sqrt{q})\sum_{y=0}^{q-1} \zeta^{xy}\ket{y}.$
If $x \ge q$, again $H_q$ leaves $\ket{x}$ unchanged. Since the resulting
transformation is a direct sum of unit matrices and matrices of the form
of $H_q$ as it was originally set down, the result is a unitary transformation.
$M_q$ and $F_q$ can be defined to operate similarly on $m$-qubit registers for
any $m$: Break up the $m$ bits into blocks of $\lceil \log q\rceil$ bits.
If $m$ is not divisible by $\lceil \log q \rceil$, then $M_q$ and $F_q$ do not
affect the ``remainder" block that contains fewer than $\lceil \log q \rceil$
bits. Likewise, in a quantum register $\ket{x_1,...,x_n}$ where each of the
$x_i$'s (with the possible exception of $x_n$)
are $\lceil \log q \rceil$-bit numbers, $M_q$
and
$F_q$ operate on the blocks of bits $x_1,...,x_n$ exactly as expected, except that 
there is no affect on the ``non-qudigit" blocks (in which $x_i \ge q$), or on
the (possibly) one remainder block for which $|x_n| < \lceil \log q \rceil$.
Since $M_q$ and $F_q$ operate exactly as they did originally on blocks
representing qudigits, and like unity for non-qudigit or remainder blocks, it is clear
that they remain unitary.

Henceforth, $M_q$, $F_q$, and $H_q$ should be understood to act on qubit
registers as described above. Nevertheless, it will usually be convenient to
think of them as acting on qu{\it digit} registers consisting of $\lceil\log
q\rceil$ qubits in each.

\begin{lemma}\label{F_q=M_q}
 $F_q$ and $M_q$ are QAC$^{(0)}$-equivalent.
\end{lemma}
\begin{proof}
By Barenco et al.~\cite{barenco95}, any fixed dimension unitary matrix can be computed 
in fixed
depth using one-qubit gates and controlled nots. Hence $H_q$ can be
computed in QAC$^{(0)}$, as can $H_q^{\otimes (n+1)}$. The
result now follows immediately from Proposition~\ref{conjugates}.
\end{proof}

 The classical Boolean $\Mod_q$-function on $n$ bits is defined so that
$\Mod_q(x_1,...,x_n) = 1$ ${\rm iff}$ $\sum_{i=1}^n x_i \equiv 0 \pmod{q}.$
 (The
more common definition sets it to 1 if $\sum_{i=1}^n x_i$ is {\it not} divisible
by $q$, but this convention is less convenient in this setting, and is not
important technically either.) We also define $\Mod_{q,r}(x_1,...,x_n)$ to
output 1 iff $\sum_{i=1}^n x_i \equiv r \pmod{q}$. Note that $\Mod_q
= \Mod_{q,0}$. Reversible, quantum versions of these functions can also be
defined. The operator $\MOD_{q,r}$ on $n+1$ qubits has the following effect:
$$\ket{x_1,...,x_n,b} \mapsto
\ket{x_1,...,x_n,b\oplus\Mod_{q,r}(x_1,...,x_n)}.$$

We write $\MOD_{q,0}$ as $\MOD_q$.
Since negation is built into the output (via the exclusive OR), it is easy to
simulate negations using $\MOD_{q,r}$ gates. For example, by setting $b=1$, we
can compute $\neg\Mod_{q,r}$. More generally, using one auxiliary bit, it is
possible to simulate ``$\neg\MOD_{q,r}$," defined so that,
$$\ket{x_1,...,x_n,b} \mapsto
\ket{x_1,...,x_n,b\oplus(\neg\Mod_{q,r}(x_1,...,x_n))},$$
using just $\MOD_{q,r}$ and a controlled-NOT gate. Thus $\MOD_{q,r}$ and
$\neg\MOD_{q,r}$ are QAC$^{(0)}$-equivalent. Moore's version of $\MOD_q$
is our $\neg\MOD_q$. Observe that $\MOD^{-1}_{q,r} = \MOD_{q,r}$.

\begin{lemma}
  $\MOD_q$ and $M_q$ are
QAC$^{(0)}$-equivalent.
\end{lemma}
\begin{proof}
  First note that $\MOD_q$ and $\MOD_{q,r}$ are equivalent, since a
$\MOD_{q,r}$ gate can be simulated by a $\MOD_q$ gate with $q-r$ extra inputs
set to the constant 1. Hence we can freely use $\MOD_{q,r}$ gates in place of
$\MOD_q$ gates.

 It is easy to see that, given an $M_q$ gate, we can simulate a $\MOD_q$ gate.
Applying $M_q$ to $n+1$ digits (represented as bits, but each digit only taking
on the values 0 or 1) transforms,
$$\ket{x_1,...,x_n,0} \mapsto
\ket{x_1,...,x_n,(\sum_i x_i)\bmod q}.$$
 Now send the bits of the last block
($\sum_i x_i\bmod q$) to a Toffoli gate with all inputs negated and 
control bit $b$. The
resulting output is exactly $b \oplus \Mod_q(x_1,...,x_n)$. The bits
in the last block 
can be erased by re-negating them and  reversing the $M_q$ gate.
This leaves only
$x_1,...,x_n$,  $O(n)$ auxiliary bits, and the output $b \oplus \Mod_q(x_1,...,x_n)$.

  The converse (simulating $M_q$ given $\MOD_q$) requires some more work. The
first step is to show that $\MOD_q$ can also determine if a sum of {\it digits}
is divisible by $q$. Let $x_1,...,x_n \in D$ be a set of digits represented as
$\lceil \log q \rceil$ bits each. For each $i$, let $x_i^{(k)}$ ($0 \le k \le
\lceil \log q \rceil - 1$) denote the bits of $x_i$.
Since the numerical value of
$x_i$ is $\sum_{k=0}^{\lceil \log q \rceil - 1} x_i^{(k)}2^k$, it follows
that
\begin{eqnarray*}
  \sum\limits_{i=1}^n x_i = \sum \limits_{k=0}^{\lceil \log q \rceil - 1}
     \sum\limits_{i=1}^n 
           x_i^{(k)}2^k.
\end{eqnarray*}

  The idea is to express this last sum in terms of a set of Boolean inputs that
are fed into a $\MOD_q$ gate. To account for the factors $2^k$, each
$x_i^{(k)}$ is fanned out $2^k$ times before plugging it into the $\MOD_q$ gate. Since 
$k < \lceil \log q \rceil$, this requires only constant depth and $O(n)$
auxiliary bits (which of course are set back to 0 in the end by reversing the
fanout). Thus, just using
$\MOD_q$ and constant fanout, we can determine if $\sum_{i=1}^n x_i \equiv
0\pmod{q}$. More generally, we can determine if $\sum_{i=1}^n x_i \equiv
r\pmod{q}$ using just a $\MOD_{q,r}$ gate and constant fanout. 
Let $\widehat{\MOD}_{q,r}(x_1,...,x_n)$ denote the resulting circuit, that
determines if a sum of digits is congruent to $r$ mod $q$.
The construction of $\widehat{\MOD}_{q,r}(x_1,...,x_n)$
is illustrated in the figure below for the case of $q=3$.
In the figure, ${\rm mod}(x)$ denotes $\Mod_{3,r}(x_1,...,x_n)$.
The notation on the right will be used as a shorthand for
this circuit:

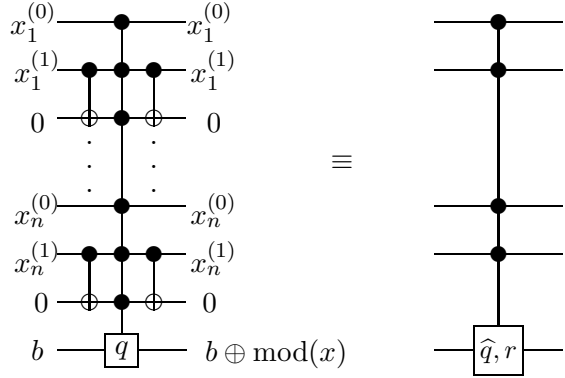
\begin{center}
\begin{picture}(3550,2800)
%\put(1300,100){\makebox(400,250)[]{Fig. 3: A $\widehat{\rm MOD}_{3,r}$ circuit for $r=0$}}
%\put(1300,-70){\makebox(400,250)[]{~~~~~~(mod$(x)$ denotes $\Mod_q(x_1,...,x_n)$).}}
\put(3050.25,2650.25){\circle*{100}}
\put(3050.25,2350.25){\circle*{100}}
\put(3050.25,1500.25){\circle*{100}}
\put(3050.25,1200.25){\circle*{100}}
%\put(2950,500){\framebox(200,200)}
\put(3000,550){\makebox(100,100)[]{\small{$\widehat{q},r$}}}
%\put(3050,2650){\line(0,-1){1950}}
\put(2650,2650){\line(1,0){800}}
\put(2650,2350){\line(1,0){800}}
\put(2650,1500){\line(1,0){800}}
\put(2650,1200){\line(1,0){800}}
\put(2650,600){\line(1,0){250}}
\put(3200,600){\line(1,0){250}}
\put(1900,1650){\makebox(350,250)[]{$\equiv$}}
\put(700.25,2650.25){\circle*{100}}
\put(700.25,2350.25){\circle*{100}}
\put(700.25,2050.25){\circle*{100}}
\put(700.25,1500.25){\circle*{100}}
\put(700.25,1200.25){\circle*{100}}
\put(700.25,900.25){\circle*{100}}
\put(600,500){\framebox(200,200)}
\put(650,550){\makebox(100,100)[]{$q$}}
\put(300,2650){\line(1,0){800}}
\put(300,2350){\line(1,0){800}}
\put(300,2050){\line(1,0){800}}
\put(500,2350){\line(0,-1){350}}
\put(900,2350){\line(0,-1){350}}
\put(500.25,2050.25){\circle{100}}
\put(900.25,2050.25){\circle{100}}
\put(300,1500){\line(1,0){800}}
\put(300,1200){\line(1,0){800}}
\put(300,900){\line(1,0){800}}
\put(700,2650){\line(0,-1){1950}}
\put(500,1200){\line(0,-1){350}}
\put(900,1200){\line(0,-1){350}}
\put(500.25,900.25){\circle{100}}
\put(900.25,900.25){\circle{100}}
\put(450,1850){\makebox(100,100)[]{.}}
\put(450,1700){\makebox(100,100)[]{.}}
\put(450,1550){\makebox(100,100)[]{.}}
\put(850,1850){\makebox(100,100)[]{.}}
\put(850,1700){\makebox(100,100)[]{.}}
\put(850,1550){\makebox(100,100)[]{.}}
\put(100,2600){\makebox(100,100)[]{$x_1^{(0)}$}}
\put(100,2300){\makebox(150,100)[]{$x_1^{(1)}$}}
\put(100,1400){\makebox(150,150)[]{$x_n^{(0)}$}}
\put(100,1950){\makebox(150,150)[]{0}}
\put(100,1100){\makebox(150,150)[]{$x_n^{(1)}$}}
\put(150,850){\makebox(100,100)[]{0}}
\put(300,600){\line(1,0){300}}
\put(800,600){\line(1,0){300}}
\put(100,550){\makebox(150,100)[]{$b$}}
\put(1200,2600){\makebox(100,100)[]{$x_1^{(0)}$}}
\put(1200,2300){\makebox(150,100)[]{$x_1^{(1)}$}}
\put(1200,1950){\makebox(150,150)[]{0}}
\put(1200,1400){\makebox(150,150)[]{$x_n^{(0)}$}}
\put(1200,1100){\makebox(150,150)[]{$x_n^{(1)}$}}
\put(1200,850){\makebox(100,100)[]{0}}
\put(1200,500){\makebox(930,160)[]{$b\oplus {\rm mod}(x)$}}
\put(500.25,2350.25){\circle*{100}}
\put(900.25,2350.25){\circle*{100}}
\put(500.25,1200.25){\circle*{100}}
\put(900.25,1200.25){\circle*{100}}
\put(2900,450){\framebox(300,300)}
\put(2900,600){\line(-1,0){250}}
\put(3200,600){\line(1,0){250}}
\put(3050,2650){\line(0,-1){1900}}
\end{picture}
\end{center}
%\caption{ A $\widehat{\rm MOD}_{3,r}$ circuit for $r=0$.
%(In the figure, ${\rm mod}(x)$ denotes $\Mod_{3,r}(x_1,...,x_n)$).}

  We can get the bits in the value of the sum $\sum_{i=1}^n x_i\bmod
q$ using $\widehat{\MOD}_{q,r}$ circuits. This is done, essentially, by
implementing the relation $x \bmod q = \sum_{r=0}^{q-1}r\cdot \Mod_{q,r}(x)$.
For each $r$, $0 \le r
\le q-1$, we compute
$\Mod_{q,r}(x_1,...,x_n)$ (where now the $x_i$'s are digits). This can
be done by applying the $\widehat{\MOD}_{q,r}$ circuits in series 
(for each $r$) to
the same inputs, introducing an auxiliary 0-bit for each application,
as illustrated here.

%\begin{figure}[h]
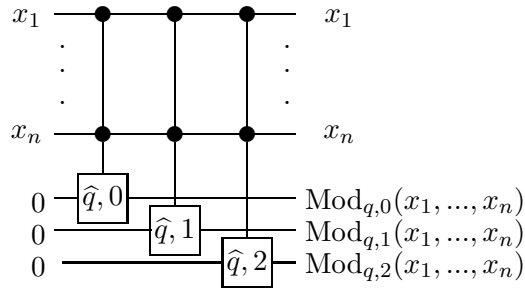
\begin{center}
\begin{picture}(2700,1950)
\put(650.25,1800.25){\circle*{100}}
\put(650.25,1050.25){\circle*{100}}
\put(1100.25,1800.25){\circle*{100}}
\put(1100.25,1050.25){\circle*{100}}
\put(1100,1750){\line(0,-1){700}}
\put(1550.25,1800.25){\circle*{100}}
\put(1550.25,1050.25){\circle*{100}}
\put(1550,1750){\line(0,-1){700}}
\put(600,600){\makebox(100,100)[]{$\widehat{q},0$}}
\put(1050,400){\makebox(100,100)[]{$\widehat{q},1$}}
\put(1500,200){\makebox(100,100)[]{$\widehat{q},2$}}
\put(350,1800){\line(1,0){1500}}
\put(350,1050){\line(1,0){1500}}
\put(2300,550){\makebox(600,150)[]{${\rm Mod}_{q,0}(x_1,...,x_n)$}}
\put(2300,350){\makebox(600,150)[]{${\rm Mod}_{q,1}(x_1,...,x_n)$}}
\put(2300,150){\makebox(600,150)[]{${\rm Mod}_{q,2}(x_1,...,x_n)$}}
\put(200,550){\makebox(100,150)[]{0}}
\put(200,350){\makebox(100,150)[]{0}}
\put(200,150){\makebox(100,150)[]{0}}
\put(100,1700){\makebox(150,150)[]{$x_1$}}
\put(2050,1700){\makebox(150,150)[]{$x_1$}}
\put(100,950){\makebox(150,200)[]{$x_n$}}
\put(2050,950){\makebox(150,200)[]{$x_n$}}
\put(350,1550){\makebox(100,100)[]{.}}
\put(350,1400){\makebox(100,100)[]{.}}
\put(350,1200){\makebox(100,100)[]{.}}
\put(1750,1550){\makebox(100,100)[]{.}}
\put(1750,1400){\makebox(100,100)[]{.}}
\put(1750,1200){\makebox(100,100)[]{.}}
\put(650,1750){\line(0,-1){650}}
\put(500,500){\framebox(300,300)}
\put(950,300){\framebox(300,300)}
\put(1400,100){\framebox(300,300)}
\put(650,1000){\line(0,-1){200}}
\put(1100,1000){\line(0,-1){400}}
\put(1550,1000){\line(0,-1){600}}
\put(500,650){\line(-1,0){150}}
\put(950,450){\line(-1,0){600}}
\put(1400,250){\line(-1,0){1000}}
\put(800,650){\line(1,0){1050}}
\put(1700,250){\line(1,0){150}}
\put(1250,450){\line(1,0){600}}
\end{picture}
\end{center}
%\caption{Applying $\widehat{{\rm MOD}}_{q,r}$ circuits in series.}
%\end{figure}

Let $r_k$ denote the $k^{th}$ bit of $r$. 
For each $r$ and for each $k$, we take the AND of the output of the
$\widehat{\MOD}_{q,r}$ with
$r_k$ (again by applying the AND's in series, which
is still constant depth, but introduces $q$ extra auxiliary inputs). Let
$a_{k,r}$ denote the output of one of these AND's. For each
$k$,
we OR together all the $a_{k,r}$'s, that is, compute
$\vee_{r=0}^{q-1} a_{k,r}$, again introducing a constant number of
auxiliary bits. Since only
one of the $r$'s will give a non-zero output from $\widehat{\MOD}_{q,r}$, this
collection of OR gates outputs exactly the bits in the value of
$\sum_{i=1}^n x_i \bmod q$. Call the resulting circuit $C$,
and the sum it outputs $S$.

 Finally, to simulate $M_q$, we need to include the input digit $b \in D$. To
do this, we apply a unitary transformation $T$ to $\ket{S, b}$ that transforms
it to $\ket{S, (b + S)\bmod q}$. 
By Barenco, et al.~\cite{barenco95}
(as in the proof of Lemma \ref{F_q=M_q}), 
$T$ can be computed in fixed depth using
one-qubit gates and controlled NOT gates.
Now using $S$ and all the other
auxiliary inputs, we reverse the computation of the circuit $C$, thus
clearing the auxiliary inputs. This is illustrated
in this figure:

%\begin{figure}[h]
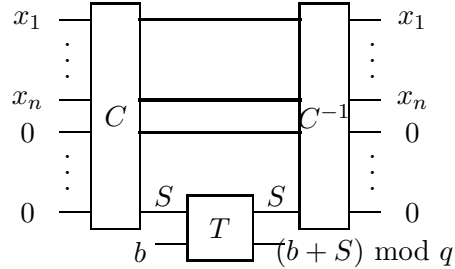
\begin{center}
\begin{picture}(2800,1850)
\put(1200,150){\framebox(400,400)}
\put(400,450){\line(1,0){200}}
\put(900,450){\line(1,0){300}}
\put(1600,450){\line(1,0){300}}
\put(2200,450){\line(1,0){200}}
\put(400,950){\line(1,0){200}}
\put(400,1150){\line(1,0){200}}
\put(400,1650){\line(1,0){200}}
\put(2200,950){\line(1,0){200}}
\put(2200,1150){\line(1,0){200}}
\put(2200,1650){\line(1,0){200}}
\put(600,350){\framebox(300,1400)}
\put(1900,350){\framebox(300,1400)}
\put(900,1650){\framebox(1000,0)}
\put(900,1150){\framebox(1000,0)}
\put(900,950){\framebox(1000,0)}
\put(100,1550){\makebox(200,200)[]{$x_1$}}
\put(100,1050){\makebox(200,200)[]{$x_n$}}
\put(2500,1550){\makebox(200,200)[]{$x_1$}}
\put(2500,1050){\makebox(200,200)[]{$x_n$}}
\put(2300,1450){\makebox(100,100)[]{.}}
\put(2300,1350){\makebox(100,100)[]{.}}
\put(2300,1250){\makebox(100,100)[]{.}}
\put(400,1450){\makebox(100,100)[]{.}}
\put(400,1350){\makebox(100,100)[]{.}}
\put(400,1250){\makebox(100,100)[]{.}}
\put(400,750){\makebox(100,100)[]{.}}
\put(400,650){\makebox(100,100)[]{.}}
\put(400,550){\makebox(100,100)[]{.}}
\put(2300,750){\makebox(100,100)[]{.}}
\put(2300,650){\makebox(100,100)[]{.}}
\put(2300,550){\makebox(100,100)[]{.}}
\put(100,850){\makebox(200,200)[]{0}}
\put(100,350){\makebox(200,200)[]{0}}
\put(2500,850){\makebox(200,200)[]{0}}
\put(2500,350){\makebox(200,200)[]{0}}
\put(950,450){\makebox(200,200)[]{$S$}}
\put(1650,450){\makebox(200,200)[]{$S$}}
\put(1300,250){\makebox(200,200)[]{$T$}}
\put(600,950){\makebox(300,200)[]{$C$}}
\put(1900,950){\makebox(300,200)[]{$C^{-1}$}}
\put(1200,250){\line(-1,0){200}}
\put(800,100){\makebox(200,200)[]{$b$}}
\put(1600,250){\line(1,0){200}}
\put(1900,100){\makebox(800,200)[]{$(b+S)~{\rm mod}~q$}}
\end{picture}
\end{center}
%\caption{Combining circuits to compute $M_q$.}
%\end{figure}

The result is an output consisting of
$x_1,...,x_n$, $O(n)$ auxiliary bits, and 
$(b + \sum_{i=1}^n x_i)\bmod q$, which is the output of an $M_q$ gate.
\end{proof}

 It is clear that we can fan out digits, and therefore bits, using an $F_q$ gate
(setting $x_i = 0$ for $1 \le i \le n$ fans out $n$ copies of $b$). It is
slightly less obvious (but still straightforward) that, given an $F_q$ gate, we
can fully simulate an $F$ gate.

\begin{lemma}\label{F-F_q-equiv}
 For any $q > 2$, $F$ and $F_q$ are QAC$^{(0)}$-equivalent.
\end{lemma}
\begin{proof} 
  By the preceeding lemmas, $F_q$ and $\MOD_q$ are QAC$^{(0)}$-equivalent.
By Moore's result, $\MOD_q$ is QAC$^{(0)}$-reducible to $F$. Hence
$F_q$ is QAC$^{(0)}$-reducible to $F$.

  Conversely, arrange each block of $\lceil \log q \rceil$ input bits to
an $F_q$ gate as follows. For the control-bit block (which contains the
bit we want to fan out), set all but the last bit to zero, and call the last
bit $b$.
Set all bits in the $i^{th}$ input-bit block to 0.
Now the $i^{th}$ output of the $F_q$ circuit
is $b$, represented as $\lceil \log q \rceil$ bits with only one possibly
nonzero bit. Send this last output bit $b$
and the input bit $x_i$ to a controlled-NOT gate.
The outputs of that gate are $b$ and $b \oplus x_i$. 
Now apply $F^{-1}_q$ to the bits that were the outputs of the $F_q$ gate
(which are all left unchanged by the controlled-not's).
This returns all the $b$'s to 0 except for
the control bit which is always unchanged. The outputs of the
controlled-not's give the desired $b \oplus x_i$.
Thus the resulting circuit simulates
$F$, with $O(n)$ auxiliary bits.
\end{proof}

\begin{theorem}
  For any $q \in \nums$, $q \not = 1$, QACC $=$ QACC[$q$].
\end{theorem}
\begin{proof}
  By the preceeding lemmas, fanout of bits is equivalent to the $\MOD_q$
function. By Moore's result, we can do $\MOD_q$ if we can do fanout in constant
depth. By our result, we can do fanout, and hence $\MOD_2$, if we can do
$\MOD_q$. Hence QACC $=$ QACC[$2$] $ \subseteq $ QACC[$q$].
\end{proof}

\section{Upper Bounds}
In this section, we prove the following upper bounds results \NQACCG
$\subseteq$ TC$^{(0)}$,
 \BQACCG $\subseteq$ TC$^{(0)}$, \NQACCP $\subseteq$
P/poly, and
 \BQACCP $\subseteq$ P/poly. 

Suppose $\{F_n\}$ and $\{z_n\}$ determine a language $L$ in NQACC. Let 
$F_n$ be the 
product of the layers $U_1, \ldots, U_t$ and $E$ be the distinct entries of
the matrices used in the $U_j$'s. By our definition of QACC, the size of 
$E$
is fixed with respect to $n$. We need a canonical way to write
sums and products of elements  in 
$E$ to be able to check $|\bra{\vec{z}}U_1\cdots U_t\ket{x,0^{p(n)}}|^2>0$
with a TC$^{(0)}$ function. To do this let $A = \{ \alpha_i\}_{1\leq i\leq 
m}$ 
be a maximal algebraically independent subset of $E$. Let $F=\rats(A)$ and 
let 
$B = \{\beta_i\}_{0\leq i < d}$ be a basis for the field $G$ generated by
the elements in $(E - A)\cup\{1\}$ over $F$.
Since the size of the bases of $F$ and $G$ are less than the
cardinality of $E$ the size of these bases is also fixed with respect 
to $n$.

As any sum or product of elements in $E$ is in $G$, it suffices
to come up with a canonical form for elements in $G$. Our representation 
is based 
on Yamakami and Yao~\cite{yy98}. Let $\alpha\in G$. Since $B$ is a basis, 
$\alpha =\sum^{d-1}_{j=0} \lambda_j
\beta_j$ for some $\lambda_j \in F$. We encode an $\alpha$ as a
$d$-tuple (we iterate the pairing  function from
the preliminaries to make $d$-tuples) 
$\langle \GN{\lambda_0}, \ldots, \GN{\lambda_{d-1}} \rangle$ where 
$\GN{\lambda_j}$
encodes $\lambda_j$. As the elements of $A$ are  algebraically 
independent, each $\lambda_j=s_j/u_j$ where $s_j$ and $u_j$ are of the
form  $$\sum_{\vec{k}_j,|\vec{k}_j| \leq e} a_{\vec{k}_j}
(\prod^m_{i=1}\alpha_i^{k_{ij}}).$$
Here $\vec{k}_j = (k_{1j},\ldots, k_{mj}) \in \ints^m$,
$|\vec{k}_j|$ is  $\sum_ik_{ij}$, 
$a_{\vec{k}_j}\in \ints$, and $e \in \nats$. In particular,
any product $\beta_m\cdot\beta_l=\sum^{d-1}_{j=0} \lambda_j \beta_j$ with
$\lambda_j = s_j/u_j$ and  $s_j$ 
and $u_j$ in this form.  
We take a common denominator $u$ for elements 
of  $E\cup\{\beta_m\cdot\beta_l\}$ and not just $E$ since the 
$\lambda_j$'s associated with the $\beta_m\cdot\beta_l$
might have additional factors in their denominators not in
$E$.
Also fix an $e$ large enough to bound the $|\vec{k_j}|$'s 
which might appear in any element of $E$ or a product 
$\beta_m\cdot\beta_l$.
This $e$ will be constant with  respect  to $n$. In multiplying $t$ layers
of QACC circuit against an input, the entries in the result
will be polynomial sums and products of elements in 
$E\cup\{\beta_m\cdot\beta_l\}$, so we can bound $|\vec{k}_j|$ for 
$\vec{k_j}$'s
which appear in the $\lambda_j$'s of such an entry by $e\cdot p(n)$.
To complete our representation of $\alpha\in G$ we encode 
$\lambda_j$ as the sequence $\langle r, \langle\langle a_{\vec{k_j}}, 
k_{1j},
\ldots, k_{mj} \rangle\rangle\rangle$ where $r$ is the power to which $u$ 
is 
raised and $\langle\langle a_{\vec{k_j}}, k_{1j},
\ldots, k_{mj} \rangle\rangle$ is the sequence of 
$\langle a_{\vec{k_j}}, k_{1j},
\ldots, k_{mj} \rangle$'s that appear in $s_j$. By our discussion,
the encoding of an $\alpha$ that appears as an entry in the output 
after applying a QACC operator to the input is of polynomial length 
and so can be manipulated in TC$^{(0)}$. 

We have need of the following lemma:

\begin{lem}
\label{sumprod}
Let $p$ be a polynomial.
(1) Let $f(i,x) \in$ TC$^{(0)}$ output encodings of $a_{i,x}\in\ints[A]$. 
Then $\ints[A]$ encodings 
of $\sum_{i=1}^{p(|x|)}a_{i,x}$ and $\prod_{i=1}^{p(|x|)}a_{i,x}$
are TC$^{(0)}$ computable.
(2) Let $f(i,x)\in$ TC$^{(0)}$ output encodings of $a_{i,x}\in G$. Then
$G$ encodings of $\sum_{i=1}^{p(|x|)}a_{i,x}$ 
and $\prod_{i=1}^{p(|x|)}a_{i,x}$
are TC$^{(0)}$ computable.
\end{lem}
\begin{proof}
We will abuse notation in this proof and identify the encoding $f(i,x)$ with
its value $a_{i,x}$. So $\sum_i f(i,x)$ and $\prod_i f(i,x)$ will mean the
encoding of $\sum_i a_{i,x}$ and $\prod_i a_{i,x}$ respectively.

(1) To do sums, the first thing we do is form the list  $L1=\langle f(0,x),
\ldots, f(p(|x|),x)\rangle$. Then we create a flattened list $L2$ from
this
 with elements  which are the $\langle a_{\vec{k_j}}, k_{1j},
\ldots, k_{mj} \rangle$'s from the $f(i,x)$'s. $L1$ is in TC$^{(0)}$ using
our definition of sequence from the preliminaries, 
and closure under sums and
$max_i$ to find the length of the longest $f(i,x)$. To flatten $L1$ we
use $max_i$ to find the length $d$ of the longest $f(i,x)$ for $i\leq
p(|x|)$. Then using max twice we can find the length of the longest
$\langle a_{\vec{k_j}}, k_{1j}, \ldots, k_{mj} \rangle$. This will be the
second coordinate in the pair used to define sequence $L2$. We then do a sum
of size $d\cdot p(|x|)$ over the subentries of $L1$ to get the first
coordinate of the pair used to define $L2$. Given $L2$, we make a list $L3$ of
the distinct $\vec{k_j}$'s that appear as  $\langle a_{\vec{k_j}}, k_{1j},
\ldots, k_{mj} \rangle$ in some $f(i,x)$ for some $i \leq p(|x|)$. This list
can be made from $L2$ using sums, $cond$ and $\mu$. We sum over the $t \leq
length(L2)$ and check if there is some $t'<t$ such that the $t'$th element of
$L2$ has same $\vec{k}_j$ as $t$ and if not add the $t$th elements $\vec{k_j}$
times 2 raised to the appropriate power. We know what power by computing the
sum of the number of smaller $t'$ that passed this test. Using $cond$ and
closure under sums we can compute in TC$^{(0)}$ a function which takes a list
like $L2$ and a $\vec{k_j}$ and returns the sum of all the $a_{\vec{k_j}}$'s
in this list. So using this function and the lists $L2$ and $L3$ we can
compute the desired encoding.

For products, since the $\alpha_i$'s of $A$ are
algebraically independent, $\ints[A]$ is isomorphic to the polynomial ring
$\ints[y_1,\ldots,y_m]$ under the  natural map which takes $\alpha_j$ to $y_j$.
We view our encodings $f(i,x)$ as $m$-variate polynomials in
$\ints[y_1,\ldots,y_m]$. We describe for any $p'$ a circuit that works
for any TC$^{(0)}$ computable $f(i,x)$ such that $\prod_i f(i,x)$ is of degree
less than $p'$ viewed as an $m$-variate polynomial. In $TC^{(0)}$ we define
$g(i,x)$ to consist of the sequence of polynomially many integer values
which result from evaluating the polynomial encoded by $f(i,x)$ at the points
$(i_1,\ldots, i_m)\in\nats^m$ where $0 \leq i_s$ and $\sum_s i_s \leq p'$. To
compute $f(i,x)$ at a point involves computing a  polynomial sum of a
polynomial product of integers, and so will be in $TC^{(0)}$. Using closure under
polynomial integer products we compute  $k(j,x) :=\prod_i \beta(j, g(i,x))$
where $\beta$ is the sequence  projection function from the preliminaries. Our
choice of points is what is called by Chung and  Yao~\cite{chungyao77} the {\em
$p'$-th order principal lattice}  of the $m$-simplex given by the origin and
the points $p'$ from the origin in each coordinate axis.  By Theorems~1 and 4
of that paper (proved  earlier by a harder argument in
Nicolaides~\cite{nic72}) the multivariate Lagrange Interpolant of degree $p'$
through the points $k(j,x)$ is unique. This interpolant is of the form
$P(y_1,\ldots,y_m) = \sum_j p_j(y_1,\ldots,y_m)k(j,x)$ where the $p_j$'s are
polynomials which do not depend on the function $f$. An explicit formula for
these $p_j$'s is given in Corollary~2 of Chung and  Yao~\cite{chungyao77} as a
polynomial product of linear factors. Since these polynomials are all of
degree less than $p'$, they have only polynomial in $p'$ many coefficients and
in PTIME these coefficients can be computed by iteratively multiplying the
linear factors together. We can then hard code these $p_j$'s (since they
don't depend on $f$) into our circuit and with these $p_j$'s, $k(j,x)$, and
closure under sums we can compute the polynomial of the desired product in
TC$^{(0)}$.

(2) We do sums first. Assume $f(i,x)
:=\sum^{d-1}_{j=0}\lambda_{ij}\beta_j$.
One immediate problem is that the $\lambda_{ij}$ and $\lambda_{i'j}$
might use different $u^r$'s for their denominators.
Since TC$^{(0)}$ is closed under
poly-sized maximum, it can find the maximum value $r_0$
to which $u$ is raised. Then it can define a function $g(i,x) =
\sum^{d-1}_{j=0}\gamma_{ij}\beta_j$
which encodes the same element of $G$ as $f(i,x)$ but where the
denominators of
the $\gamma_{ij}$'s are now $u^{r_0}$. If $\lambda_j$ was $s_j/u^r$ we need
to compute the encoding $s_j\cdot u^{r_0-r}/u^{r_0}$. This is
straightforward
from (1). Now
$$\sum_{i=1}^{p(|x|)}f(i,x) = \sum_{i=1}^{p(|x|)}g(i,x) =
\sum^{d-1}_{j=0}[(\sum_{i=1}^{p(|x|)}s_{ij})/u^{r_0}]\beta_j,$$
where $s_{ij}$'s are the numerators of the $\gamma_{ij}$'s in $g(i,x)$.
From part (1) we can compute the encoding $e_j$ of
$(\sum_{i=1}^{p(|x|)}s_{ij})$
in TC$^{(0)}$.  So the desired answer
$\langle \langle r_0, e_0\rangle, \cdots, \langle r_0, e_{d-1}\rangle
\rangle$ is in TC$^{(0)}$.

For products $\prod_{i=1}^{p(|x|)}f(i,x)$, we play the same trick as the 
in the $\ints[A]$ product case. We view our encodings of elements of $G$ as
d-variate polynomials in $F(y_0,\ldots,y_{d-1})$ under the map $\beta_k$ goes
to $y_k$. (Note that this map is not necessarily an isomorphism.) We then create a
function $g(i,x)$ which consists of the sequence of values obtained by
evaluating $f(i,x)$ at polynomially many points in a lattice as in the first
part of this lemma. Evaluating $f(i,x)$ at a point can easily be done using
the first part of this lemma. We then use part (1) of this lemma to compute
the products $k(j,x) = \beta(j,g(i,x))$. We then get the interpolant $P(y_0,
\ldots, y_{d-1}) = \sum_j p_j(y_0,\ldots,y_m)k(j,x)$.  We non-uniformly obtain
the encoding of  $p_j(\beta_0,\ldots, \beta_{d-1})$ expressed
as an element of $G$. i.e., in the form $\sum^{d-1}_{w=0} \lambda_{jw}\beta_w$.
Thus, the product $\prod_{i=1}^{p(|x|)}f(i,x)$ is
$$\sum^{d-1}_{w=0}(\sum_j\lambda_{jw}k(j,w)) \beta_w$$
The encoding of the products is the d-tuple given by 
$\langle \sum_j\lambda_{j0}k(j,0),\ldots,\sum_j\lambda_{jd-1}k(j,d-1) \rangle$
Each of its components is a
polynomial sum of a product of two things in $F$ and can be
computed using the first part of the lemma.
\end{proof}

  For $\{F_n\}\in$ QAC$^{(0)}_{wf}=$ QACC, the vectors that $F_n$  act on
are elements of a 
$2^{n+p(n)}$ dimensional space $\mathcal{E}$$_{1,n+p(n)}$ space 
which
is a tensor product of the 2-dimensional spaces 
$\mathcal{E}$$_1, \ldots \mathcal{E}$$_{n+p(n)}$, which in turn are each
spanned by $\ket{0}, \ket{1}$. We write $\mathcal{E}$$_{j,k}$ for the 
subspace  
$\otimes^{k}_{i=j} \mathcal{E}$$_i$ of $\mathcal{E}$$_{1,n+p(n)}$.  We now
define a succinct way to represent a set of vectors in 
$\mathcal{E}$$_{1,n+p(n)}$ which is useful in our argument below. A {\em
tensor graph} is a directed acyclic graph with one source node of indegree
zero,  one terminal node of outdegree zero, and two kinds of edges: 
horizontal
edges, which are unlabeled, and vertical edges, which are labeled with
a pair of amplitudes and a product of {\em colors} and {\em anticolors}. (The
color product may be
the number 1.)  We require that
all paths  from the source to the terminal traverse the same number of 
vertical
edges and that no vertex can have vertical edge indegree greater than one 
or
outdegree greater than one. For a color $c$ we
write $\tilde{c}$ for its corresponding anticolor. The {\em height} of a 
node 
in a tensor
graph is the number of vertical edges traversed to get to it on any path 
from
the source; the {\em height} of an edge is the height of its end node.
The {\em width} of a tensor graph is maximum number of nodes of the same 
height. As an example of a
tensor graph where our color product is the number 1, consider the 
following
figure:

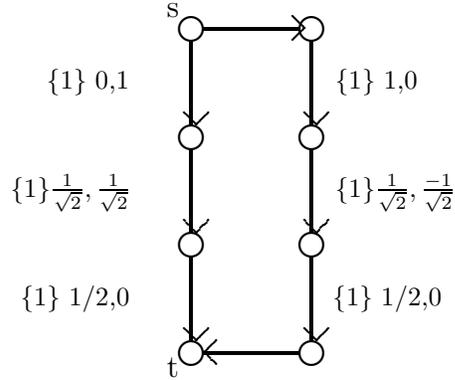
\begin{center}
\begin{picture}(2025,2343)(2101,-2707)
\thicklines
\put(3226,-511){\circle{150}}
\put(3226,-1186){\circle{150}}
\put(3226,-1861){\circle{150}}
\put(3226,-2536){\circle{150}}
\put(3226,-586){\line( 0,-1){525}}
\put(3226,-1261){\line( 0,-1){525}}
\put(3226,-1936){\line( 0,-1){525}}
\put(3977,-511){\circle{150}}
\put(3977,-1186){\circle{150}}
\put(3977,-1861){\circle{150}}
\put(3977,-2536){\circle{150}}
\put(3977,-586){\line( 0,-1){525}}
\put(3977,-1261){\line( 0,-1){525}}
\put(3977,-1936){\line( 0,-1){525}}
\put(3301,-2536){\line( 1, 0){600}}
\put(3301,-511){\line( 1, 0){600}}
\multiput(3151,-1036)(6.25000,-6.25000){13}{\makebox(6.6667,10.0000){
\SetFigFont{7}{8.4}{rm}.}}
\multiput(3226,-1111)(6.25000,6.25000){13}{\makebox(6.6667,10.0000){
\SetFigFont{7}{8.4}{rm}.}}
\multiput(3826,-436)(6.25000,-6.25000){13}{\makebox(6.6667,10.0000){
\SetFigFont{7}{8.4}{rm}.}}
\multiput(3901,-511)(-6.25000,-6.25000){13}{\makebox(6.6667,10.0000){
\SetFigFont{7}{8.4}{rm}.}}
\multiput(3901,-1036)(6.25000,-6.25000){13}{\makebox(6.6667,10.0000){
\SetFigFont{7}{8.4}{rm}.}}
\multiput(3976,-1111)(6.25000,6.25000){13}{\makebox(6.6667,10.0000){
\SetFigFont{7}{8.4}{rm}.}}
\multiput(3151,-1711)(6.25000,-6.25000){13}{\makebox(6.6667,10.0000){
\SetFigFont{7}{8.4}{rm}.}}
\multiput(3226,-1786)(6.25000,6.25000){13}{\makebox(6.6667,10.0000){
\SetFigFont{7}{8.4}{rm}.}}
\multiput(3151,-2386)(6.25000,-6.25000){13}{\makebox(6.6667,10.0000){
\SetFigFont{7}{8.4}{rm}.}}
\multiput(3226,-2461)(6.25000,6.25000){13}{\makebox(6.6667,10.0000){
\SetFigFont{7}{8.4}{rm}.}}
\multiput(3901,-1711)(6.25000,-6.25000){13}{\makebox(6.6667,10.0000){
\SetFigFont{7}{8.4}{rm}.}}
\multiput(3976,-1786)(6.25000,6.25000){13}{\makebox(6.6667,10.0000){
\SetFigFont{7}{8.4}{rm}.}}
\multiput(3901,-2386)(6.25000,-6.25000){13}{\makebox(6.6667,10.0000){
\SetFigFont{7}{8.4}{rm}.}}
\multiput(3976,-2461)(6.25000,6.25000){13}{\makebox(6.6667,10.0000){
\SetFigFont{7}{8.4}{rm}.}}

\multiput(3351,-2611)(-6.25000,6.25000){13}{\makebox(6.6667,10.0000){
\SetFigFont{7}{8.4}{rm}.}}
\multiput(3276,-2536)(6.25000,6.25000){13}{\makebox(6.6667,10.0000){
\SetFigFont{7}{8.4}{rm}.}}

\put(3076,-436){\makebox(0,0)[lb]{\smash{\SetFigFont{12}{14.4}{rm}s}}}
\put(3076,-2686){\makebox(0,0)[lb]{\smash{\SetFigFont{12}{14.4}{rm}t}}}
\put(2326,-886){\makebox(0,0)[lb]{\smash{\SetFigFont{10}{16.8}{rm}\{1\} 
0,1}}}
\put(2101,-1561){\makebox(0,0)[lb]{\smash{\SetFigFont{10}{16.8}{rm}$\{1\}
 \frac{1}{\sqrt{2}},\frac{1}{\sqrt{2}}$}}}
%\put(2301,-1561){\makebox(0,0)[lb]{\smash{\SetFigFont{10}{16.8}{rm}$\{1\}
% \frac{1}{\sqrt{2}},\frac{1}{\sqrt{2}}$}}}
\put(2101,-2236){\makebox(0,0)[lb]{\smash{
\SetFigFont{10}{16.8}{rm}\{1\} 1/2,0}}}
%\put(2326,-886){\makebox(0,0)[lb]{\smash{\SetFigFont{10}{16.8}{rm}\{1\} 
%0,1}}}
%\put(2326,-886){\makebox(0,0)[lb]{\smash{\SetFigFont{10}{16.8}{rm}\{1\} 
%0,1}}}
\put(4126,-886){\makebox(0,0)[lb]{\smash{\SetFigFont{10}{16.8}{rm}\{1\} 
1,0}}}
\put(4126,-1561){\makebox(0,0)[lb]{\smash{\SetFigFont{10}{16.8}{rm}$\{1\}
\frac{1}{\sqrt{2}},\frac{-1}{\sqrt{2}}$}}}
\put(4051,-2236){\makebox(0,0)[lb]{\smash{
\SetFigFont{10}{16.8}{rm}\{1\} 1/2,0}}}
\end{picture}
\end{center}
The rough idea of tensor graphs is that paths through the graph
correspond to collections of vector in $\mathcal{E}_{1,n}$. For this 
particular
figure the left path from the source node (s) to the terminal node (t)
corresponds to the vectors given by 
$$\ket{1}\otimes (\frac{1}{\sqrt{2}}\ket{0} +
\frac{1}{\sqrt{2}}\ket{1})\otimes \frac{1}{2}\ket{0}$$
and the right hand path corresponds to
$$\ket{0}\otimes (\frac{1}{\sqrt{2}}\ket{0} +
\frac{-1}{\sqrt{2}}\ket{1})\otimes \frac{1}{2}\ket{0}.$$

A $\mathcal{E}$$_{j,k}$-{\em term} in a tensor graph is a maximal
induced tensor subgraph between a node of height $j-1$ and a node of height
$k$. We also require that the horizontal indegree of the node at height 
$j-1$
be zero and that the horizontal outdegree of the node at height $k$ be 
zero.
For the graph we considered above there are two $\mathcal{E}$$_{1,2}$-terms
and two $\mathcal{E}$$_{2,3}$-terms but only one $\mathcal{E}$$_{1,3}$-term
corresponding to the whole figure.

Colors are used to handle controlled-not layers. A color $c$ and
its anticolor $\tilde{c}$ satisfy the following multiplicative properties:
$c\cdot c =\tilde{c} \cdot \tilde{c} = 1$ and $c \cdot \tilde{c} = 0$.
Given two distinct colors $b$ and $c$ we have $b \cdot c = c \cdot b$ and
$\tilde{b} \cdot c = c \cdot \tilde{b}$. If $a$ is a product of colors and
anticolors not involving the color $b$ or $\tilde{b}$ and $c$
is another product of colors we have $a(bc)=(ab)c$. 
We consider formal sums of products of complex numbers times colors.
We require  complex numbers to commute with colors and require colors and
anticolors to distribute, i.e., if $a$, $b$, $c$ are colors or anticolors
then $a\cdot (b+c) = a\cdot b + a\cdot c$ and $(b+c)\cdot a = b\cdot a + c
\cdot a$. Finally, we require addition to work so that the above structure
satisfies the axioms of an $\complexes$-algebra.
Given a tensor graph $G$ denote by 
$\mathcal{A}_{G}$ the $\complexes$-algebra above.
Since 
$$(a\cdot a)\cdot \tilde{a} = \tilde{a} \neq 0=a\cdot(a \cdot\tilde{a})$$
this algebra is not associative. However, in the sums we will consider
the terms will never have more than two positions where a color or its
anticolor can occur, so the products we will consider are associative. 
Using our our earlier encoding for the elements of
$\complexes$ which could appear in a $QACC$ computation,
it is straightforward to use sequence coding to get a TC$^{(0)}$ encodings of 
the relevant elements of $\mathcal{A}_G$. As an example of how
colors affect amplitudes, consider the  following picture:

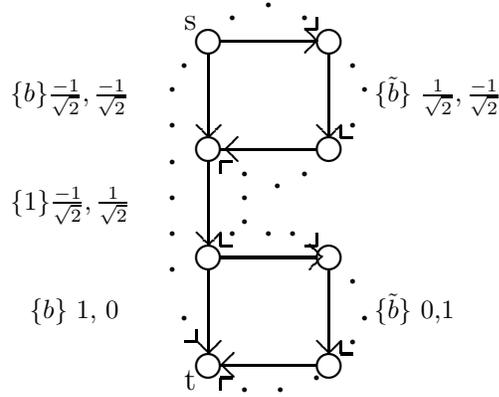
\begin{center}
\begin{picture}(2133,2466)(2101,-2719)
\thicklines
\put(3226,-511){\circle{150}}
\put(3226,-1186){\circle{150}}
\put(3226,-1861){\circle{150}}
\put(3226,-2536){\circle{150}}
\put(3226,-586){\line( 0,-1){525}}
\put(3226,-1261){\line( 0,-1){525}}
\put(3226,-1936){\line( 0,-1){525}}
\put(3977,-511){\circle{150}}
\put(3977,-1186){\circle{150}}
\put(3977,-1861){\circle{150}}
\put(3977,-2536){\circle{150}}
\put(3376,-361){\circle{20}}
\put(3601,-286){\circle{20}}
\put(3826,-361){\circle{20}}
\put(4126,-661){\circle{20}}
\put(4201,-811){\circle{20}}
\put(4126,-1036){\circle{20}}
\put(3451,-1336){\circle{20}}
\put(3826,-1336){\circle{20}}
\put(3656,-1411){\circle{20}}
\put(3451,-1486){\circle{20}}
\put(3376,-1711){\circle{20}}
\put(3451,-1636){\circle{20}}
\put(3076,-661){\circle{20}}
\put(3001,-811){\circle{20}}
\put(3001,-1036){\circle{20}}
\put(3001,-1261){\circle{20}}
\put(3001,-1486){\circle{20}}
\put(3001,-1711){\circle{20}}
\put(3001,-1936){\circle{20}}
\put(3076,-2236){\circle{20}}
\put(3751,-1711){\circle{20}}
\put(3601,-1711){\circle{20}}
\put(4126,-2011){\circle{20}}
\put(4201,-2236){\circle{20}}
\put(4126,-2386){\circle{20}}
\put(3451,-2686){\circle{20}}
\put(3676,-2686){\circle{20}}
\put(3901,-2611){\circle{20}}
\put(3301,-2536){\line( 1, 0){600}}
\put(3301,-511){\line( 1, 0){600}}
\multiput(3151,-1036)(6.25000,-6.25000){13}{\makebox(6.6667,10.0000)
{\SetFigFont{7}{8.4}{rm}.}}
\multiput(3226,-1111)(6.25000,6.25000){13}{\makebox(6.6667,10.0000)
{\SetFigFont{7}{8.4}{rm}.}}
\multiput(3826,-436)(6.25000,-6.25000){13}{\makebox(6.6667,10.0000)
{\SetFigFont{7}{8.4}{rm}.}}
\multiput(3901,-511)(-6.25000,-6.25000){13}{\makebox(6.6667,10.0000)
{\SetFigFont{7}{8.4}{rm}.}}
\multiput(3901,-1036)(6.25000,-6.25000){13}{\makebox(6.6667,10.0000)
{\SetFigFont{7}{8.4}{rm}.}}
\multiput(3976,-1111)(6.25000,6.25000){13}{\makebox(6.6667,10.0000)
{\SetFigFont{7}{8.4}{rm}.}}
\multiput(3151,-1711)(6.25000,-6.25000){13}{\makebox(6.6667,10.0000)
{\SetFigFont{7}{8.4}{rm}.}}
\multiput(3226,-1786)(6.25000,6.25000){13}{\makebox(6.6667,10.0000)
{\SetFigFont{7}{8.4}{rm}.}}
\multiput(3151,-2386)(6.25000,-6.25000){13}{\makebox(6.6667,10.0000)
{\SetFigFont{7}{8.4}{rm}.}}
\multiput(3226,-2461)(6.25000,6.25000){13}{\makebox(6.6667,10.0000)
{\SetFigFont{7}{8.4}{rm}.}}
\multiput(3901,-2386)(6.25000,-6.25000){13}{\makebox(6.6667,10.0000)
{\SetFigFont{7}{8.4}{rm}.}}
\multiput(3976,-2461)(6.25000,6.25000){13}{\makebox(6.6667,10.0000)
{\SetFigFont{7}{8.4}{rm}.}}

\multiput(3351,-2611)(-6.25000,6.25000){13}{\makebox(6.6667,10.0000){
\SetFigFont{7}{8.4}{rm}.}}
\multiput(3276,-2536)(6.25000,6.25000){13}{\makebox(6.6667,10.0000){
\SetFigFont{7}{8.4}{rm}.}}
\multiput(3376,-1261)(-6.25000,6.25000){13}{\makebox(6.6667,10.0000){
\SetFigFont{7}{8.4}{rm}.}}
\multiput(3301,-1186)(6.25000,6.25000){13}{\makebox(6.6667,10.0000){
\SetFigFont{7}{8.4}{rm}.}}
\multiput(3826,-1786)(6.25000,-6.25000){13}{\makebox(6.6667,10.0000){
\SetFigFont{7}{8.4}{rm}.}}
\multiput(3901,-1861)(-6.25000,-6.25000){13}{\makebox(6.6667,10.0000){
\SetFigFont{7}{8.4}{rm}.}}

\put(3977,-586){\line( 0,-1){525}}
\put(3977,-1936){\line( 0,-1){525}}
\put(3901,-1186){\line(-1, 0){600}}
\put(3901,-1861){\line(-1, 0){600}}
\put(3301,-2611){\line( 0, 1){  0}}
\put(3301,-2611){\line( 1, 0){ 75}}
\put(3301,-2686){\line( 0, 1){ 75}}
\put(4051,-2386){\line( 0,-1){ 75}}
\put(4051,-2461){\line( 1, 0){ 75}}
\put(3826,-1786){\line( 1, 0){ 75}}
\put(3901,-1786){\line( 0, 1){ 75}}
\put(3826,-436){\line( 1, 0){ 75}}
\put(3901,-436){\line( 0, 1){ 75}}
\put(4051,-1036){\line( 0,-1){ 75}}
\put(4051,-1111){\line( 1, 0){ 75}}
\put(3301,-1261){\line( 0, 1){  0}}
\put(3301,-1261){\line( 1, 0){ 75}}
\put(3301,-1261){\line( 0,-1){ 75}}
\put(3301,-1711){\line( 0,-1){ 75}}
\put(3301,-1786){\line( 1, 0){ 75}}
\put(3151,-2311){\line( 0,-1){ 75}}
\put(3151,-2386){\line(-1, 0){ 75}}
\put(3076,-436){\makebox(0,0)[lb]{\smash{\SetFigFont{12}{14.4}{rm}s}}}
\put(3076,-2686){\makebox(0,0)[lb]{\smash{\SetFigFont{12}{14.4}{rm}t}}}
\put(1931,-886){\makebox(0,0)[lb]{\smash{
\SetFigFont{10}{16.8}{rm}\{$b$\}$\frac{-1}{\sqrt{2}},\frac{-1}{\sqrt{2}}$}}}
\put(1931,-1561){\makebox(0,0)[lb]{\smash{
\SetFigFont{10}{16.8}{rm}\{1\}$\frac{-1}{\sqrt{2}},\frac{1}{\sqrt{2}}$}}}
\put(2051,-2236){\makebox(0,0)[lb]{\smash{
%\SetFigFont{10}{16.8}{rm}\{b\} 1,0}}}
%\put(4200,-886){\makebox(0,0)[lb]{\smash{
\SetFigFont{10}{16.8}{rm}\{$b$\} 1, 0}}}
\put(4200,-886){\makebox(0,0)[lb]{\smash{
\SetFigFont{10}{16.8}{rm}\{$\tilde{b}$\} $\frac{1}{\sqrt{2}},\frac{-1}{\sqrt{2}}$}}}
\put(4200,-2236){\makebox(0,0)[lb]{\smash{
\SetFigFont{10}{16.8}{rm}\{$\tilde{b}$\} 0,1}}}
\end{picture}
\end{center}
The amplitude of $\ket{1,0,0}$ in the left hand dotted path is
$b \cdot \frac{-1}{\sqrt{2}} \cdot 1 \cdot \frac{-1}{\sqrt{2}} \cdot b
\cdot 1 =
1/2$ using commutativity and $b^2=1$. Its amplitude in the right hand dotted
path would be zero because of the last vertical edge. However, vectors such
as $\ket{0,0,1}$ would have nonzero amplitude in the right hand dotted path.
Nevertheless, the amplitude of any vector $\ket{\vec{x}}$ in any path other than the
dotted ones from $s$ to $t$ will be $0$ as $b\cdot\tilde{b}=0$. More
formally, we define the amplitude of an $\ket{\vec{x}}$ in a vertical edge as equal
to the left amplitude times the color product in the edge if $\vec{x}$ is
$\ket{0}$ and equal to the right amplitude times the color product in the edge if
$\vec{x}$ is $\vec{1}$.  The amplitude of a vector $\ket{x_1,\ldots,x_j}$ in a
path in a tensor graph is the product over $k$ from 1 to $j$ of the amplitude
of the vectors $\ket{x_k}$ in the vertical edge of height $k$. The amplitude of a 
vector $\ket{x_j,\ldots, x_k}$
in an $\mathcal{E}_{j,k}$-term is the sum of its amplitude in its paths. The
amplitude of a vector $\ket{x_1,\ldots,x_{p(n)}}$ in a tensor graph $G$ is  defined
to be the sum of its amplitudes in $G$'s $\mathcal{E}_{1,p(n)}$-terms. 

As we will be interested in families of tensor graphs $\{G_n\}$, corresponding
to our circuit families we want to look at those families with a certain degree
of uniformity. We say a family of tensor graphs $\{G_n\}$ is {\em color consistent}
if:
(1) the number of colors for edges of the same height is bounded by a constant $k$ 
with respect to $n$, (2) the number of heights in which a given
color/anticolor can appear is exactly two (colors and their anticolors must
appear on the same heights), (3) each color product at the same height is of
the form $\prod^k_{i=0} l_i$ where $l_i$ must be either a color $c_i$ or
$\tilde{c_i}$ (it follows there are $2^k$ possible color products for edges at
 a given height). We say that a color/anticolor is {\em active} at a given
height if the height is at or after the first height at which the color/anticolor 
occurs and is below the height of its second occurrence. The family is
further said to be {\em log-color depth } if the number of active 
colors/anticolors of a given height is log-bounded.
%By {\em adding a term to a
%tensor graph} we mean adding a horizontal edge out of some node in the graph
%into the source node of a term followed by a horizontal edge from the terminal
%of the into some other node in the graph. 

\begin{thm}
\label{multlayers}
Let $\{F_n\}$ be a family of QACC operators and let $\{\bra{\vec{z}_n}\}$ a
family of 
 observables. (1) There is a color-consistent family of tensor
graphs of width
 $2^{2^{2t}}$ and polynomial size representing the output
amplitudes of
 $U_1\cdots U_t \ket{\vec{z}_n}$ where $U_i$ are the layers of
$F_n$. 
 (2) If $\{F_n\}$ is in \QACCP then the family of tensor graphs will
be of
 log-color depth. (3) If $\{F_n\}$ is in \QACCG 
then the number of paths from the source to the terminal node is polynomially
bounded.
\end{thm} 
\begin{proof}
The proof is by induction on $t$. In the base case, $t=0$, we do not
multiply any layers, and we can easily represent this as a tensor
graph of width 1. Assume for $j <t$ that 
$U_j\cdots U_1\ket{\vec{x}, 0^{p(n)}}$ can be written as 
color consistent tensor graph of width $2^{2^{2t}}$ and polynomial size. There
are two cases to consider: In the first case the layer is a tensor  product of
matrices $M_1 \otimes \cdots \otimes M_\nu$ where the  $M_k$'s are Toffoli
gates, one qubit gates, or fan-out gates (since QAC$^{(0)}_{wf}=$QACC); in the
second case the layer is a controlled-not layer. 

For the first case we ``multiply'' $U_t$ against our current graph by 
``multiplying'' each $M_j$ in parallel against the terms in our sum 
corresponding to $M_j$'s domain, say $\mathcal{E}$$_{j',k'}$. If  $M_j = \left(
\begin{array}{cc} u_{00} & u_{01} \\ u_{10} &  u_{11}\end{array} \right)$ with
domain $\mathcal{E}$$_{j'}$ is a one-qubit gate, then we multiply the two 
amplitudes in each vertical edge of height $j'$ in our tensor graph by $M_j$. 
This does not effect the width, size, or number of paths through the graph.  If
$M_j$ is a Toffoli gate, then for each term $S$ in  $\mathcal{E}$$_{j',k'}$ in
our tensor graph we add one new term to the resulting graph. This term is
added by adding a horizontal edge going out from the source node of $S$
followed by the new $\mathcal{E}$$_{j',k'}$-term followed by a horizontal edge
into the terminal node of $S$. The new term is obtained from the old one
 by setting
to
$0$ the left hand amplitudes of all edges in $S$ of height between $j'$ and $k'-1$
and then if $\alpha,\gamma$ is the amplitude of an edge of height $k'$ in the
new term we change it to $\gamma -\alpha, \alpha-\gamma$. This new term
adjusts the amplitude for the case of a $\ket{1}^{\otimes(k'-j'-1)}$ vector
in $\mathcal{E}$$_{j',k'-1}$ tensored with either a $\ket{0}$ or $\ket{1}$. 
This operation increases the width of the new tensor graph by the width of
the $\mathcal{E}_{j',k'}$-term for
each $\mathcal{E}$$_{j',k'}$-term in the graph. Since the original graph has
width $2^{2^{2(t-1)}}$ there are at most this many starting and ending
vertices for such terms.  So there at most $(2^{2^{2(t-1)}})^2$ such terms.
Each of these terms has width at most
$2^{2^{2(t-1)}}$.
Thus, the new width is at most
$$2^{2^{2(t-1)}} + (2^{2^{2(t-1)}})^2\cdot 2^{2^{2(t-1)}}< 2^{2^{2t}}.$$  
Notice this action adds one new path through the $\mathcal{E}_{j',k'}$ part of
the graph for every existing one.

Now suppose $M_j$ is a
fan-out gate, let  $S$ be a $\mathcal{E}$$_{j',k'}$-term in our tensor graph
and let $e$ be any vertical edge in $S$ in $\mathcal{E}$$_{k'}$. Suppose $e$
has amplitude $\alpha$  for $\ket{0}$ and amplitude $\gamma$ for $\ket{1}$. In
the new graph we change the amplitude of $e$ to $\alpha,0$. We then add a
horizontal edge out of the source node of $S$ followed by a new
$\mathcal{E}$$_{j',k'}$-term followed by  a horizontal  edge into the terminal
node of $S$. The new term is obtained from $S$ by changing the amplitude for
edges in $\mathcal{E}_{k'}$ with amplitudes $\alpha,\gamma$ in $S$ to $0,\gamma$.
The amplitudes of the non-$\mathcal{E}_{k'}$ edges in this term are the
reverse of the corresponding edge in $S$, i.e., if the edge in $S$ had
amplitude $\delta,\zeta$ then the new term edge would have amplitude
$\zeta,\delta$. The same argument as in the Toffoli case shows the new width
is bounded by $2^{2^{2t}}$ and that this action adds one new path through the
$\mathcal{E}_{j',k'}$ part of
the graph for every existing one.

For the case of a controlled-not layer, suppose we have a
controlled-not going from line $i$ onto line $j$.
Let $c, \bar{c}$ be a new color, anti-color pair not yet appearing in the
graph. Let $e_i$ be a vertical edge of height $i$ in the graph and let 
$C_i, \alpha_i,\gamma_i$ be respectively its color product and two amplitudes.
Similarly, let $e_j$ be a vertical edge of height $j$  in the graph 
and $C_j, \alpha_j,\gamma_j$ be its color product and two
amplitudes. In the new graph we multiply $c$ times the color product of $e_i$ and
$e_j$ and change the amplitude of $e_i$ to $\alpha_i,0$. We then add a
horizontal edge going out from the starting node of $e_i$, followed by a
vertical edge with values $C_i\cdot\tilde{c}, 0, \gamma_i$ followed by a
horizontal edge into the terminal node of $e_i$. In turn, we add a horizontal edge
going out of the starting node of $e_j$, followed by a vertical edge with
values $C_j\cdot\tilde{c}, \gamma_i, \alpha_j$ followed by a horizontal edge into
the terminal node of $e_j$.  We handle all other controlled gates in this
layer in a similar fashion (recall they must go to disjoint lines). 
We add at most a new vertex of a given height for every existing vertex of a
given height. So the total
width is at most doubled by this operation and $2\cdot2^{2^{2(t-1)}} <
2^{2^{2t}}$. In the \QACCP case, simulating a layer which is a Kronecker product 
of spaced controlled-not gates and identity matrices, notice we would at most
add one to the color depth at any place. So if a controlled-not layer is
a composition of $O(\log)$ many such layers it will increase the color depth
by $O(\log)$. In the \QACCG \ case, notice that simulating a single controlled-not
we add one new path for each existing path through the graph at each of
the two heights affected. This gives three new paths on the whole subspace
for each old one.

Since we have handled the two possible layer cases and the changes we needed
to make only increase the resulting tensor graph polynomially, we
thus have established the induction step and (1) and (2) of the theorem.
For (3), observe for each multi-line gate we handle in adding a layer
we at most quadruple the number of paths through the subspace where that gate
applies. Since there are at most logarithmically many such gates, the number
of paths through the graph increases polynomially.
\end{proof}

\begin{theorem}
\label{amplitudes}
Let $\{G_n\}$ be a family of constant width color-consistent tensor graphs
of vectors in 
$\mathcal{E}$$_{1,p(n)}$. Assume the coefficients of amplitudes in the
$\{G_n\}$ can be encoded in $TC^{(0)}$ using our encoding scheme
described earlier and that $\{G_n\}$ has log-color depth. 
Then the amplitude of any basis vector of $\mathcal{E}$$_{1,p(n)}$ in $G_n$ 
is P/poly computable. If the number of paths through the graph from the source
to the terminal node is polynomially bounded then the amplitude of any basis
vector is TC$^{(0)}$ computable. 
\end{theorem}
\begin{proof}
Let $G_n$ be a particular graph in the family and let $\ket{\vec{x_n}}$ be the vector
whose amplitude we want to compute.
Assume that all graphs in our family
have fewer than $k$ colors in any color product and have a width bounded by $w$.  
We will proceed from the source to the terminal node one height at a time to
compute the amplitude. Since the width is $w$ the number of
$\mathcal{E}_1$-terms is at most $w$ and each of these must have width
at most $w$. Let $\alpha_{1,1}, \ldots, \alpha_{1,w}$  (some of which may be zero)
denote the amplitudes in $\mathcal{A}_{G_n}$ of $\ket{x_{n,1}}$ in each of
these terms. The $\alpha_{1,i}$ are each sums of at most $w$ amplitudes times
the color products of at most $k$ colors and anticolors, so the encoding of
these $w$ amplitudes is TC$^{(0)}$ computable. Because of the restriction on
the width of $G_n$ there are at most $w$ many $\mathcal{E}$$_{1,j}$-terms,
$w^2$ many $\mathcal{E}$$_{j,j+1}$-terms,  and $w$ many
$\mathcal{E}$$_{1,j+1}$-terms. Fixing some ordering on the nodes of height $j$
and $j+1$ let $\gamma_{j,i,k}$ be the amplitude of $\ket{x_{n,j+1}}$ in the
$\mathcal{E}$$_{j,j+1}$-term with source the $i$th node of height $j$ and with
terminal node the $k$th node of height $j+1$. The amplitude is zero if there
is no such $\mathcal{E}_{j,j+1}$-term. Then the amplitudes $\alpha_{j+1,1},
\ldots, \alpha_{j+1,w}$ of the $\mathcal{E}$$_{1,j+1}$-terms can be computed
from the amplitudes $\alpha_{j,1}, \ldots, \alpha_{j,w}$ of the
$\mathcal{E}$$_{1,j}$-terms using the formula
$$\alpha_{j+1,k} = \sum_{i=1}^w \alpha_{j,i}\cdot \gamma_{j,i,k}.$$ 
Thus $\alpha_{j+1,k}$ can be computed from the $\alpha_{j,i}$ using
a polynomial sized circuit to do these adds and multiplies. Similarly,
each $\alpha_{j,k}$ can be computed by polynomial sized circuits from the
$\alpha_{j-1,k}$'s and so on. Since we have log-color depth the number of
terms consisting of elements in our field times color products in a
$\alpha_{j,k}$ will be polynomial. So the size of the $\alpha_{j,k}$'s $j \leq
p(n)$, $k \leq w$ will be polynomial in the input $\vec{x}_n$. So the size of
the circuits for each $\alpha_{j,k}$ where $j\leq p(n)$ and $k\leq w$ will be
polynomial size. There is only one $\mathcal{E}$$_{1,p(n)}$-term in $G_n$ and
its amplitude is that of $\ket{\vec{x}_n}$, so this shows it has polynomial
sized circuits.  

For the TC$^{(0)}$ result, if the number of paths is polynomially bounded, then
the amplitude can be written as the polynomial sum of the amplitudes in each
path. The amplitude in a path can in turn be calculated as a polynomial
product of the amplitudes times the colors on the vertical edges in the path.
Our condition on every color appearing at exactly two heights guarantees the
color product along the whole path will be 1 or 0, and will be zero iff we get
a color and its anticolor on the path. This is straightforward to check in
TC$^{(0)}$, so this sum of products can thus be computed in TC$^{(0)}$ using
Lemma~\ref{sumprod}.
\end{proof}

\begin{cor}
\item[ (1)] \EQACCP$\subseteq$\NQACCP$\subseteq$P/Poly, and  
\BQACCP\\ $\subseteq$P/poly.
\item[ (2)] \EQACCG$\subseteq$\NQACCG$\subseteq$TC$^{(0)}$, and\\ 
\BQACCG$ \subseteq$TC$^{(0)}$.
 
\end{cor}
\begin{proof}
Given a a family $\{F_n\}$ of \QACCP operators and a family
$\{\bra{\vec{z}_n}\}$ of states we can use Theorem~\ref{multlayers} to get a
family $\{G_n\}$ of log color depth, color-consistent tensor graphs
representing the amplitudes of $F^{-1}_n\ket{\vec{z}_n}$. Note $\{F^{-1}_n\}$
is also a family of \QACCP operators since Toffoli and fan-out gates are their
own inverses, the inverse of any one qubit gate is also a one qubit gate
(albeit usually a different one), and finally a controlled-not layer is its own
inverse. Theorem~\ref{amplitudes} shows there is a P/poly circuit
computing the  amplitude of any vector $\ket{\vec{x}_n}$ in this graph. This
amounts to calculating 
$$\bra{\vec{x}_n} F^{-1}_n \ket{\vec{z}_n} = \bra{\vec{z}_n} F_n \ket{\vec{x}_n}$$
If this is nonzero, then
$|\bra{\vec{z}_n} F_n \ket{\vec{x}_n}|^2 >0$, 
and we know $\vec{x}$ is in the language. In the BQACC$_{\rats}$ 
case everything is a rational so P/poly can explicitly 
compute the magnitude of the amplitude and check if it is greater than $3/4$.
The TC$^{(0)}$ result follows similarly from the TC$^{(0)}$ part of
Theorem~\ref{multlayers}.  
\end{proof}

\section{Discussion and Open Problems}

A number of
questions are suggested by our work. 

\begin{itemize}
\item Is all of NQACC in TC$^{(0)}$ or even P/Poly? 
We conjecture that NQACC is in TC$^{(0)}$. As mentioned
in the introduction, we have developed techniques that
remove some of the important obstacles to proving
this.
\item Are there any natural problems in NQACC
that are not known to be in
ACC?
\item What  exactly is the complexity of  the 
languages in EQACC, NQACC and BQACC$_{\rats}$? 
We entertain two extreme possibilities.
Recall that the class ACC can be computed by
quasipolynomial size depth 3 threshold circuits~\cite{yao90}. It would be 
quite
remarkable if EQACC could also be simulated in that manner. However, it is 
far from
clear if  any of the
techniques used in the simulations of ACC (the Valiant-Vazirani lemma,
composition of low-degree polynomials,
modulus amplification via the Toda polynomials, etc.), which seem to be
inherently irreversible, can be applied in the quantum setting.
At the other extreme, it would be equally remarkable if NQACC and
NQTC$^{(0)}$ (or BQACC$_{\rats}$ and NQTC$^{(0)}$) coincide. 
Unfortunately, an optimal characterization of QACC language classes
anywhere between those two extremes
would probably require new (and probably difficult) proof techniques.
\item How hard are the fixed levels of QACC? While lower bounds for QACC 
itself seem
impossible at present, it might be fruitful to study the limitations
of small depth QACC circuits (depth 2, for example).
%\item Finally, can the depth bounds on the TC$^{(0)}$ circuits to simulate
%NQACC circuits be improved? 
\end{itemize}

\noindent {\bf Acknowledgments}:
We thank Cris Moore for pointing out an error in an earlier version of
Theorem~\ref{multlayers},
and Bill Gasarch for helpful comments and suggestions.


\begin{thebibliography}{1}

\bibitem{barenco95}
A.~Barenco, C.~Bennett, R.~Cleve, D.P.~DiVincenzo, N.~Margolus, P.~Shor,
T.~Sleator, J.A.~Smolin, and H. Weifurter.
\newblock Elementary gates for quantum computation.
\newblock Phys. Rev. A {\bf 52}, pages 3457--3467, 1995.

\bibitem{chungyao77}
K.~C.~Chung and T.~H.~Yao.
\newblock On Lattices Admitting Unique Lagrange Interpolations .
\newblock Siam Journal of Numerical Analysis {\bf 14}, pages 735--743, 1977.

\bibitem{clote93}
P.~Clote.
\newblock On polynomial Size Frege Proofs of Certain Combinatorial Principles.
\newblock In P.~Clote and J.~Krajicek, editors, {\em Arithmetic, Proof
Theory, and Computational Complexity},   pages 164--184. Oxford, 1993.

\bibitem{fj99}
L.~Fortnow and J. Rogers.
\newblock Complexity Limitations on Quantum Computation.
\newblock {\em Proceedings of 13th IEEE Conference on Computational 
Complexity}, pages 202--209, 1998.

\bibitem{fghr98}
S.~Fenner, F.~Green, S.~Homer, and R.~Pruim.
\newblock Quantum NP is hard for PH.
\newblock {\em Proceedings of 6th Italian Conference on theoretical 
Computer Science}, World Scientific, Singapore, pages 241--252, 1998. 

\bibitem{maciel98}
Alexis Maciel and Denis Therien.
\newblock Threshold Circuits of Small Majority-Depth.
\newblock Information and Computation {\bf 146}. 55--83, 1998.

\bibitem{moore98}
Cristopher Moore and Martin Nilsson.
\newblock Parallel Quantum Computation and Quantum Codes
\newblock In Los Alamos Preprint archives (1998), quant-ph/9808027.

\bibitem{moore99}
Cristopher Moore.
\newblock Quantum Circuits: Fanout, Parity, and Counting.
\newblock In Los Alamos Preprint archives (1999), quant-ph/9903046.

\bibitem{nic72}
R.~A.~Nicolaides.
\newblock On a class of finite elements generated by Lagrange interpolation.
\newblock Siam Journal of Numerical Analysis {\bf 9}, pages 177--199, 1972.


\bibitem{shor97}
P.~W. Shor.
\newblock  Polynomial-time algorithms for prime number factorization and
discrete logarithms on a quantum computer.
\newblock  {\em SIAM J. Comp.}, 26:1484--1509, 1997.

\bibitem{siu91}
K.-Y.~Siu and V~Rowchowdhury.
\newblock On optimal depth threshold circuits for multiplication and 
related problems.
\newblock {\em SIAM J. Discrete Math.} {\bf 7}. 284--292, 1994.

\bibitem{siu94}
K.-Y.~Siu and J~Bruck.
\newblock On the power of threshold circuits with small weights
\newblock {\em SIAM J. Discrete Math.} {\bf 4}. 423--435, 1991.

\bibitem{smo87}
R.~Smolensky. 
\newblock Algebraic methods in the
theory of lower bounds for Boolean circuit complexity. 
\newblock {\em Proceedings of the 19th Annual ACM Symposium on Theory of
Computing.} 77-82, 1987.


\bibitem{yy98}
T. Yamakami and A.C. Yao.
\newblock {$NQP_{\bf C}= co$-$C_=P$.}
\newblock To appear in {\it Information Processing Letters}.

\bibitem{yao90}
A.~C.-C. Yao.
\newblock On ACC and threshold circuits.  \newblock In {\it Proceedings of
the 31st Symposium on Foundations of Computer Science},
(1990), 619-627.

\bibitem{yao93}
A.~C.-C. Yao.
\newblock  Quantum circuit complexity.
\newblock In {\em Proceedings of the 34th IEEE Symposium on Foundations of
Computer Science}, pages 352--361, 1993.

\end{thebibliography}
\end{document}